\documentclass[11pt]{elsarticle}

\usepackage{setspace}
\usepackage{algpseudocode}
\usepackage{mathrsfs}
\usepackage{multirow}
\usepackage{tikz}
\usepackage{setspace}
\usepackage{etoolbox}   
\usepackage{lipsum}
\usepackage{booktabs}
\usepackage{graphicx,wrapfig}

\makeatletter
\def\ps@pprintTitle{%
  \let\@oddhead\@empty
  \let\@evenhead\@empty
  \def\@oddfoot{\reset@font\hfil\thepage\hfil}
  \let\@evenfoot\@oddfoot
}
\makeatother

\AtBeginEnvironment{tabular}{\doublespacing}

\usetikzlibrary{arrows,automata,backgrounds,fit,patterns,petri,shadows,shapes, positioning}

\usepackage[top=3.5cm,bottom=3.5cm,left=3.5cm,right=3.5cm]{geometry}

\usepackage{amsthm}

\usepackage{lineno}
\usepackage[hidelinks]{hyperref}
\modulolinenumbers[5]

\usepackage{amssymb}
\usepackage{mathtools}
\usepackage{mathrsfs}  
\usepackage{float}
\usepackage{graphicx}  

\usepackage{amsmath}
\usepackage{graphicx,psfrag,epsf}
\usepackage{enumerate}
\usepackage{natbib}
\usepackage{comment}
\usepackage[official]{eurosym}
\usepackage{eurosym}
\usepackage{url}  
\usepackage{siunitx}
\usepackage{bbm}

\usepackage{url}            
\usepackage{booktabs}      
\usepackage{subcaption}
\usepackage{amsfonts}      
\usepackage{nicefrac}       
\usepackage{microtype}      

\usepackage{amsmath}
\usepackage{mathabx}
\usepackage{amsopn}

\usepackage{algpseudocode}
\usepackage{algorithm}

\algrenewcommand\algorithmicindent{0.5em}

\makeatletter
\renewcommand{\Function}[2]{
  \csname ALG@cmd@\ALG@L @Function\endcsname{#1}{#2}
  \def\jayden@currentfunction{#1}
}

\newcommand{\funclabel}[1]{
  \@bsphack
  \protected@write\@auxout{}{
    \string\newlabel{#1}{{\jayden@currentfunction}{\thepage}}
  }
  \@esphack
}

\usepackage{color}
\usepackage{caption}

\usepackage{float}
\floatstyle{plaintop}
\restylefloat{table}

\usepackage{amsthm}
\usepackage{bm}

\DeclareMathOperator*{\argmax}{arg\,max}

\newcommand{\dd}{\mathop{}\! \mathrm{d}}

\usepackage{color,soul}
\usepackage[usestackEOL]{stackengine}[2013-09-11]

\setstackgap{L}{1.4\baselineskip}
\usepackage{tikz}
\usetikzlibrary{arrows,automata,backgrounds,fit,patterns,petri,shadows,shapes, positioning}
\pgfdeclarepatternformonly{stripes}
{\pgfpointorigin}{\pgfpoint{0.25cm}{0.25cm}}
{\pgfpoint{0.25cm}{0.25cm}}
{
    \pgfpathmoveto{\pgfpoint{0cm}{0cm}}
    \pgfpathlineto{\pgfpoint{0.25cm}{0.25cm}}
    \pgfpathlineto{\pgfpoint{0.25cm}{0.12cm}}
    \pgfpathlineto{\pgfpoint{0.12cm}{0cm}}
    \pgfpathclose%
    \pgfusepath{fill}
    \pgfpathmoveto{\pgfpoint{0cm}{0.12cm}}
    \pgfpathlineto{\pgfpoint{0cm}{0.25cm}}
    \pgfpathlineto{\pgfpoint{0.12cm}{0.25cm}}
    \pgfpathclose%
    \pgfusepath{fill}
}

\usepackage{tabularx}
\usepackage{booktabs}

\biboptions{authoryear}

\begin{document}

\begin{frontmatter}

\title{Manipulating Hidden-Markov-Model Inferences by Corrupting Batch Data}

\author[my1address]{William N. Caballero}

\author[my2address]{Jose Manuel Camacho}
\author[my3address]{Tahir Ekin}
\author[my4address]{Roi Naveiro}

\address[my1address]{United States Air Force Academy, Colorado Springs, Colorado, USA.}
\address[my2address]{Institute of Mathematical Sciences (ICMAT), Madrid, Spain.}
\address[my3address]{McCoy College of Business, Texas State University, San Marcos, Texas, USA.}
\address[my4address]{CUNEF Universidad, Madrid, Spain.}

\begin{abstract}
Time-series models typically assume untainted and legitimate streams of data. However, a self-interested adversary may have incentive to corrupt this data, thereby altering a decision maker's inference. Within the broader field of adversarial machine learning, this research provides a novel, probabilistic perspective toward the manipulation of hidden Markov model inferences via corrupted data. In particular, we provision a suite of corruption problems for filtering, smoothing, and decoding inferences leveraging an adversarial risk analysis approach. Multiple stochastic programming models are set forth that incorporate realistic uncertainties and varied attacker objectives. Three general solution methods are developed by alternatively viewing the problem from frequentist and Bayesian perspectives. The efficacy of each method is illustrated via extensive, empirical testing. The developed methods are characterized by their solution quality and computational effort, resulting in a stratification of techniques across varying problem-instance architectures. This research highlights the weaknesses of hidden Markov models under adversarial activity, thereby motivating the need for robustification techniques to ensure their security. 
\end{abstract}


\end{frontmatter}

\section{Introduction}

\begin{figure}[htbp]
\centering
\begin{tikzpicture}[ ->,>=stealth',shorten >=1pt,auto,node distance=1.5cm,
                    semithick]
    \tikzstyle{theta1}=[circle,
                                    thick,
                                    minimum size=1.15  cm,
                                    draw=black,
                                    fill=white]
    \tikzstyle{theta2}=[circle,
                                    thick,
                                    minimum size=1.15  cm,
                                    draw=black,
                                    fill=white]
    \tikzstyle{thetan}=[circle,
                                    thick,
                                    minimum size=1.15  cm,
                                    draw=black,
                                    fill=white]
    \tikzstyle{thetaempty}=[circle,
                                    thick,
                                    minimum size=1.15  cm,
                                    draw=black,
                                    fill=white]
    \tikzstyle{thetaN}=[circle,
                                    thick,
                                    minimum size=1.15  cm,
                                    draw=black,
                                    fill=white]
    \tikzstyle{X1}=[circle,
                                    thick,
                                    minimum size=1.15  cm,
                                    draw=black,
                                    fill=white]
    \tikzstyle{X2}=[circle,
                                    thick,
                                    minimum size=1.15  cm,
                                    draw=black,
                                    fill=white]
    \tikzstyle{Xn}=[circle,
                                    thick,
                                    minimum size=1.15  cm,
                                    draw=black,
                                    fill=white]
    \tikzstyle{Xempty}=[circle,
                                    thick,
                                    minimum size=1.15  cm,
                                    draw=black,
                                    fill=white]
    \tikzstyle{XN}=[circle,
                                    thick,
                                    minimum size=1.15  cm,
                                    draw=black,
                                    fill=white]
  \tikzstyle{texto}=[label]	
    \node[theta1](A)   {$Q_{1} $};
  \node[theta2](B)  [right of=A]   {$Q_{2} $};
  \node[thetan](E)  [right of=B]   {$Q_{3}$};
\node[draw=none](E1)  [right of=E]   {$...$};
  \node[thetaN](F)  [right of=E1]   {$Q_{|\mathcal{T}|} $};
  \node[X1] (C) [below of=A]  {$X_{1}$};
   \node[X2] (D) [below of=B]  {$X_{2}$};
   \node[Xn] (G) [below of=E]  {$X_{3}$};
   \node[draw=none] (G1) [below of=E1]  {$...$};
   \node[XN] (H) [below of=F]  {$X_{|\mathcal{T}|}$};
  \path (A) edge    node {} (B)
         (A) edge             node {} (C)
        (B) edge    node {} (E)
        (B) edge              node {} (D)
        (E) edge    node {} (E1)
        (E) edge              node {} (G)
        (E1) edge    node {} (F)
        (F) edge    node {} (H);
\end{tikzpicture} 
 \caption{A basic hidden Markov model} \label{fig:hmm}
\end{figure}
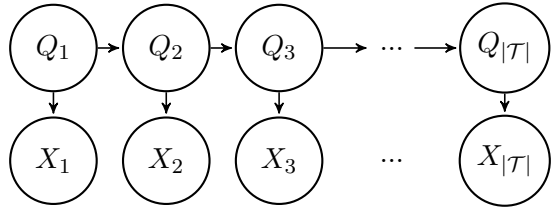

Hidden Markov models (HMMs) are stochastic processes over some time interval $\mathcal{T}$ which assume that an observed sequence of outputs, $\{x_t\}_{t\in \mathcal{T}}$, is a noisy observation from an underlying unobservable (hidden) sequence of states, $\{q_t\}_{t\in \mathcal{T}}$, that is, in turn, generated by a Markov chain, $\{Q_t\}_{t \in \mathcal{T}}$. In canonical HMMs, the state evolution model is a discrete-time, discrete-state, first-order Markov chain such that $\mathcal{T}=\{1, 2, \ldots, |\mathcal{T}|\}$ and $Q_t \in \mathcal{Q}=\{1, \ldots ,|\mathcal{Q}|\}$. Transitions between $Q_t$ and $Q_{t+1}$ are governed by the state-transition probability matrix $A$ having elements $a_{ij}=P(Q_{t+1}=j|Q_{t}=i), \forall (i,j)\in \mathcal{Q} \times \mathcal{Q}$, and initial-state probabilities are given by $\pi_i=P(Q_1=i), \forall i \in \mathcal{Q}$. Furthermore, each observation $X_t \in \mathcal{X} = \{1, \ldots |\mathcal{X}|\}$ is determined by the observation probability matrix $B$ having elements $b_{ik}=P(X_{t}=k|Q_{t}=i), \ \forall (i,k) \in \mathcal{Q} \times \mathcal{X}$. Figure \ref{fig:hmm} graphically depicts this standard model; the nodes and arcs represent random variables and conditional dependencies, respectively.

Despite their relative simplicity, HMMs can be leveraged to perform a wide array of inferences and predictions. \citet{rabiner1989tutorial} discusses three commonly considered problems: (1) identifying the 
likelihood $P( \{ X_t \}_{t\in \mathcal{T}})$ for a given HMM parameterization, (2) selecting the most likely sequence of latent states given an HMM model and a sequence of observations, i.e., decoding, and (3) learning to optimally parameterize, through maximum likelihood estimation, an HMM given a sequence of observations (e.g., estimating $A$ and $B$). In addition to these standard problems, researchers and practitioners are also often interested in the probability of a specific hidden state at some time $t$. That is, given a sequence of observations, one may wish to infer $P(Q_{|\mathcal{T}|}=i | \{X_{\tau}\}_{\tau \in \mathcal{T}} =\{x_{\tau}\}_{\tau \in \mathcal{T}})$ and $P(Q_{t}=i | \{X_{\tau}\}_{\tau \in \mathcal{T}} =\{x_{\tau}\}_{\tau \in \mathcal{T}})$ for $t<|\mathcal{T}|$. These inferences are also known as filtering and smoothing, respectively. Bayesian approaches to HMMs \citep[e.g., see][]{scott} also draw upon these building blocks, but vary mechanically based upon their alternative perspective. Moreover, whereas inference may be performed sequentially, HMMs often consider inference on batch data, i.e., using complete sequences of observations.

The utility of such predictions and inferences has been illustrated in myriad applications and diverse disciplines. From gesture recognition in computer science \citep{starner1997real} and chromatin state learning in computational biology \citep{ernst2012chromhmm} to stochastic thermodynamics in physics \citep{bechhoefer2015hidden} and signal processing in electrical engineering \citep{crouse1998wavelet}, HMMs have proven themselves to effectively characterize unobservable states based upon another related, observable process. Of such applications, speech recognition is, perhaps, the most successful and widely known historical use of HMMs \citep{gales2008application} dating back to the canonical works summarized by \citet{rabiner1989tutorial}.
Security applications are also relevant including network 
intrusion detection \citep{scott}  and spam detection \citep{conde}.

Such successful applications drive increased incorporation of HMMs into commercial products and, as discussed by \citet{biggio2018wild}, this increased utilization is accompanied by a wide array of security threats. More specifically, if an automated system utilizes a machine learning algorithm, a nefarious actor may attempt to subvert the system by manipulating its underlying statistical framework (e.g., by providing it corrupted data). The field of \textit{adversarial machine learning} (AML) focuses on modeling these threats from both an offensive and a defensive perspective, i.e., modeling how the attacks affect the algorithm and how to defend against the associated negative effects. Although AML is a relatively nascent field, much progress has been made in the last fifteen years since \citet{dalvi2004adversarial} published the discipline's seminal work regarding adversarial classification. As a result of deep learning's recent widespread adoption in commercial products, neural networks have received substantial interest in the AML literature \citep{melis2017deep, crecchi2020fader, sotgiu2020deep}, with support vector machines likely being a close second \citep{xiao2015support, indyk2019adversarial, alhajjar2021adversarial}. Although classification and computer vision remain the predominant focus of AML research, other machine learning tasks have been emphasized in more recent investigations \citep[e.g., see][]{jagielski2018manipulating, caballero2021poisoning, gallego2019reinforcement}. In particular, adversarial machine learning algorithms relating to temporal data and unsupervised learning are emerging areas of inquiry \citep[e.g., see][]{alfeld2016data, chen2020optimal, dang2020adversarial, naveiro2021adversarial, hsu2021adversarial}.

However, given the shear breadth and variety of available machine learning methods and variations, many techniques have been sparingly studied (if at all) from an AML perspective. HMMs are one such understudied technique of particular relevance \citep{CABALLERO2020}. Real-world application of the attacks abound, ranging from SMS spam misclassification \citep{xia2020discrete} and target detection \citep{miller2015hidden} to crisis prediction in international politics \citep{o2010crisis}. Therefore, to address this gap in the literature, we develop herein multiple algorithmic methodologies that can be leveraged to corrupt HMM data such that the resulting inference is erroneous and benefits the interests of a nefarious actor. The development of such attacks is paramount to ensure the security of canonical HMM algorithms when the veracity of the underlying data is threatened; their weaknesses must first be identified before they can be strengthened. 

This manuscript makes two primary contributions. Firstly, we provide a comprehensive collection of HMM corruption problems that encompass filtering, smoothing, and decoding inferences under realistic uncertainty conditions. Secondly, we develop and benchmark a set of attack frameworks that can be tailored to specific scenarios. Through extensive empirical testing, we demonstrate the effectiveness of our attacks and highlight the trade-off between solution quality and computational effort when corrupting larger-scale HMMs. In so doing, our research explores the limits of the developed frameworks, providing insight into their capabilities.

The organization for the remainder of this manuscript is as follows. We begin in Section \ref{secBack} by providing requisite background on HMM inferences and our opponent modeling framework (i.e., adversarial risk analysis, \cite{Banks:2015}). Subsequently, Section \ref{secProbs} formally defines a collection of HMM corruption problems for use against filtering, smoothing, and decoding inferences. Not only are the resultant mathematical programs non-linear and combinatorial in nature, but they are also characterized by uncertain parameters. Therefore, the considered models are stochastic programming problems for which most traditional solution techniques are unsuitable. Therefore, in Section \ref{SecSolution}, we set forth three approximation methods that enable the attacker to tractably identify high-quality solutions. These solution techniques are tested in Section \ref{secTRA} via a designed experiment over a subset of tunable, algorithmic parameters. The practical relevance of our methods is also illustrated via a case study whereby an HMM used for part-of-speech tagging is attacked. Finally, we conclude this manuscript in Section \ref{secConc} by providing closing remarks and presenting promising avenues of future research. The appendix provides implementation details of the computational algorithms.

\section{Relevant, Contextual Background} \label{secBack}

This manuscript is multi-disciplinary in nature, integrating machine-learning and decision-analytic techniques within a competitive stochastic process. To ensure comprehension by a broad readership, this section briefly summarizes relevant background material regarding HMMs and adversarial risk analysis (ARA). 

\subsection{Inference and Prediction on Hidden Markov Models} \label{secBackHMM}

HMMs, such as that provided in Figure \ref{fig:hmm}, are probabilistic graphical models utilized in myriad applications. Akin to other graphical models (e.g., Bayesian networks), it is often of interest to update probabilities in the HMM by conditioning upon a subset of known variables. Given that $Q_t$ is unobservable, such updates are performed in an HMM assuming some set of observed $\{x_t\}_{t \in \mathcal{T}}$.  Provided this \textit{batch} of observations, filtering, smoothing, and decoding are among the most frequently solved problems\footnote{The learning problem, i.e., fitting the most-likely HMM parameters given a set of observations, is another common HMM problem but not one that we address herein. We refer the interested reader to \citet{rabiner1989tutorial} for more information in this regard.}.

\small
\begin{algorithm}[!htbp]
\caption{Forward Algorithm } 
\label{algForward}
\begin{algorithmic}[]
\For{ $i \in \mathcal{Q}$}
    \State{ $\alpha_{1,i} = \pi_i b_{i, x_1}$}
\EndFor
\For{$\tau = 2, \ldots, t$ and $i \in \mathcal{Q}$}
    \State{ $\alpha_{\tau,i} = \sum_{j \in \mathcal{Q}} \alpha_{\tau-1,j} a_{j,i} b_{i, x_\tau} $}
\EndFor
\end{algorithmic}
\end{algorithm}
\normalsize

The filtering and smoothing problems can each be solved utilizing the Forward and Backward algorithms summarized in Algorithms \ref{algForward} and \ref{algBackward}. In the case of filtering, only Algorithm \ref{algForward} must be utilized to identify the forward probabilities, i.e. 

$$
\alpha_{t,i}=P\left(Q_{t} =i, \{X_{\tau}\}_{\tau=1}^t =\{x_{\tau}\}_{\tau=1}^t \right).$$ 

\noindent Once identified, it can be shown that

$$P(Q_{|\mathcal{T}|} =i
| \{X_{\tau}\}_{\tau \in \mathcal{T}} =\{x_{\tau}\}_{\tau \in \mathcal{T}}
) =\frac{\alpha_{|\mathcal{T}|,i}}{\sum_{j \in \mathcal{Q}} \alpha_{|\mathcal{T}|,j}}.$$

\small
\begin{algorithm}[!htbp]
\caption{Backward Algorithm } 
\label{algBackward}
\begin{algorithmic}[1]
\For{ $i \in \mathcal{Q}$}
    \State{ $\beta_{|\mathcal{T}|,i} = 1$}
\EndFor
\For{$\tau = |\mathcal{T}|-1, \ldots, t$ and $i \in \mathcal{Q}$}
    \State{ $\beta_{\tau,i} = \sum_{j \in \mathcal{Q}} \beta_{t+1,j} a_{i,j} b_{j, x_{\tau+1}}$}
\EndFor
\end{algorithmic}
\end{algorithm}
\normalsize

\noindent Alternatively, for smoothing problems, the time of interest is some $t < |\mathcal{T}|$, implying that both the forward and backward algorithms must be leveraged. 
Once the backward probabilities, i.e., 

$$
\beta_{t,i}=P (\{X_{\tau}\}_{\tau =t+1}^{|\mathcal{T}|} =\{x_{\tau}\}_{\tau = t+1}^{|\mathcal{T}|} | Q_t =i ).$$ 

\noindent are identified via Algorithm \ref{algBackward}, it is known that 

\begin{equation}
P\left(Q_{t} =i
| \{X_{\tau}\}_{\tau \in \mathcal{T}} =\{x_{\tau}\}_{\tau \in \mathcal{T}}\right) =\frac{\alpha_{t,i}\beta_{t,i}}{\sum_{j \in \mathcal{Q}} \alpha_{t,j}\beta_{t,j}}. \label{eqSmooth}
\end{equation}

Decoding problems, for their part, are typically addressed via application of the Viterbi algorithm summarized in Algorithm \ref{algViterbi}. Given some $\{x_t\}_{t \in \mathcal{T}}$, this algorithm functions by successively identifying the probability of the most likely latent-state path to $Q_t=i$, $\forall t \in \mathcal{T}, i\in \mathcal{Q}$. These are referred to as Viterbi probabilities and denoted by $\delta_{t,i}$. By storing the backtraces, i.e., $\psi_{t,i}$, the most likely sequences of states can be reconstructed by identifying $q_{|\mathcal{T}|}^* = \argmax_{i \in \mathcal{Q}} \delta_{|\mathcal{T}|,i}$ and tracing the corresponding chain of $\psi_{t,i}$-values back to $t=1$.

\small
\begin{algorithm}[!htbp]
\caption{Viterbi Algorithm } 
\label{algViterbi}
\begin{algorithmic}
\For{ $i \in \mathcal{Q}$}
    \State{Set $\delta_{1,i} = \pi_i b_{i, x_1}$ and $\psi_{1,i}=0$}
\EndFor
\For{$t = 2, \ldots, |\mathcal{T}|$ and $i \in \mathcal{Q}$}
    \State{$\delta_{t,i} = \max_{j \in \mathcal{Q}} \delta_{t-1,j} a_{j,i} b_{i, x_t}$ }
    \State{$\psi_{t,i} = \argmax_{j \in \mathcal{Q}} \delta_{t-1,j} a_{j,i}$ }
\EndFor
\State{Set $q^*_{|\mathcal{T}|}= \argmax_{i \in \mathcal{Q}} \delta_{|\mathcal{T}|,i}$}
\State{Set $q^*_{t}= \psi_{t+1, q^*_{t+1}}$ for $t=|\mathcal{T}|-1, ...,1$}
\end{algorithmic}
\end{algorithm}
\normalsize

\subsection{Adversarial Risk Analysis} \label{secBackARA}

ARA is a Bayesian alternative to game-theoretic analysis of a competitive interaction. Whereas game theory solves every player's problem simultaneously in a given solution concept, ARA approaches the problem from a decision-theoretic perspective. That is, ARA extends canonical, decision-analytic methods to competitive interactions, by enabling a supported player to maximize their expected utility under aleatory, epistemic, and solution-concept uncertainties. Within this research, we adopt the ARA perspective toward opponent modeling to decompose the game into a decision problem for the attacker. This is in juxtaposition to a game-theoretic approach that solves the game as a system to identify all possible equilibrium profiles.

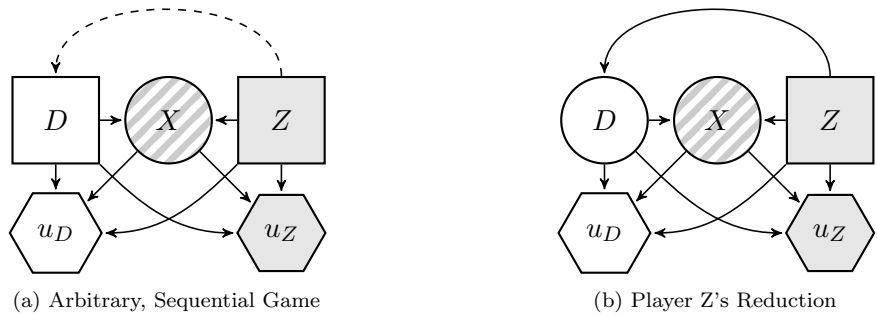
\begin{figure}[htbp!]
\centering
\begin{subfigure}{.5\textwidth}
  \centering
  \begin{tikzpicture}[->,>=stealth',shorten >=1pt,auto,node distance=1.5cm,
                    semithick]
  \tikzstyle{uncertain}=[circle,pattern=stripes,
                                    pattern color=gray!40,
                                    thick,
                                    minimum size=1.15 cm,
                                    draw=black]
  \tikzstyle{uncertainX}=[circle,pattern=stripes,
                                    pattern color=gray!40,
                                    thick,
                                    minimum size=1.15 cm,
                                    draw=black]
  \tikzstyle{uncertainY}=[circle,pattern=stripes,
                                    pattern color=gray!40,
                                    thick,
                                    minimum size=1.15 cm,
                                    draw=black]
  \tikzstyle{Attacker_utility}=[regular polygon,regular polygon sides=6,
                                    thick,
                                    minimum size=1.15 cm,
                                    draw=black,
                                    fill=gray!20]
  \tikzstyle{Defender_utility}=[regular polygon,regular polygon sides=6,
                                    thick,
                                    minimum size=1.15 cm,
                                    draw=black,
                                    fill=white]
  \tikzstyle{Attacker_decision}=[rectangle,
                                    thick,
                                    minimum size=1.15  cm,
                                    draw=black,
                                    fill=gray!20]
  \tikzstyle{Defender_decision}=[rectangle,
                                    thick,
                                    minimum size=1.15  cm,
                                    draw=black,
                                    fill=white]
  \tikzstyle{texto}=[label]			
  \node[uncertain](X)   {$X$};
  \node[Defender_decision] (D) [left of=X]  {$D$};
  \node[Attacker_decision] (Z) [right of=X] {$Z$};
  \node[Defender_utility]  (uD) [below of=D] {$u_D$};
  \node[Attacker_utility]  (uZ) [below of=Z] {$u_Z$};
  \path (D) edge    node {} (uD)
            edge    node {} (X)
            edge[out=-45, in = 180]    node {} (uZ)
        (Z) edge[out=-135, in = 0]    node {} (uD)
            edge    node {} (uZ)
            edge    node {} (X)
            edge[out=90, in =90, dashed]    node {} (D)
       (X) edge node{} (uD)
            edge  node{} (uZ);
\end{tikzpicture}
  \caption{Arbitrary, Sequential Game}
  \label{subfig:BAID1}
\end{subfigure}%
\begin{subfigure}{.5\textwidth}
  \centering
  \begin{tikzpicture}[->,>=stealth',shorten >=1pt,auto,node distance=1.5cm,
                    semithick]
  \tikzstyle{uncertain}=[circle,pattern=stripes,
                                    pattern color=gray!40,
                                    thick,
                                    minimum size=1.15 cm,
                                    draw=black]
   \tikzstyle{Defender_uncertainty}=[circle,                                   fill=white,
                            thick,
                            minimum size=1.15 cm,
                                    draw=black]
  \tikzstyle{uncertainX}=[circle,pattern=stripes,
                                    pattern color=gray!40,
                                    thick,
                                    minimum size=1.15 cm,
                                    draw=black]
  \tikzstyle{uncertainY}=[circle,pattern=stripes,
                                    pattern color=gray!40,
                                    thick,
                                    minimum size=1.15 cm,
                                    draw=black]
  \tikzstyle{Attacker_utility}=[regular polygon,regular polygon sides=6,
                                    thick,
                                    minimum size=1.15 cm,
                                    draw=black,
                                    fill=gray!20]
  \tikzstyle{Defender_utility}=[regular polygon,regular polygon sides=6,
                                    thick,
                                    minimum size=1.15 cm,
                                    draw=black,
                                    fill=white]
  \tikzstyle{Attacker_decision}=[rectangle,
                                    thick,
                                    minimum size=1.15  cm,
                                    draw=black,
                                    fill=gray!20]
  \tikzstyle{Defender_decision}=[rectangle,
                                    thick,
                                    minimum size=1.15  cm,
                                    draw=black,
                                    fill=white]
  \tikzstyle{texto}=[label]			
  \node[uncertain](X)   {$X$};
  \node[Defender_uncertainty] (D) [left of=X]  {$D$};
  \node[Attacker_decision] (Z) [right of=X] {$Z$};
  \node[Defender_utility]  (uD) [below of=D] {$u_D$};
  \node[Attacker_utility]  (uZ) [below of=Z] {$u_Z$};
  \path (D) edge    node {} (uD)
            edge    node {} (X)
            edge[out=-45, in = 180]    node {} (uZ)
        (Z) edge[out=-135, in = 0]    node {} (uD)
            edge    node {} (uZ)
            edge    node {} (X)
            edge[out=90, in =90]    node {} (D)
       (X) edge node {} (uD)
            edge node {} (uZ);;
\end{tikzpicture}
  \caption{Player Z's Reduction}
  \label{subfig:BAID2}
\end{subfigure}
\caption{A Visualization of the ARA Approach in a Sequential Game}
\label{fig:BAID}
\end{figure}

The ARA approach to a competitive interaction can be visualized via the bi-agent influence diagrams (BAID) in Figure \ref{subfig:BAID1} and the corresponding ARA reduction in Figure \ref{subfig:BAID2}. Squares represent decision nodes for the corresponding players (i.e, $D$ or $Z$), circles represent uncertainty nodes, and hexagons are utility nodes. A white fill corresponds to Player-$D$ nodes and a gray fill to Player-$Z$ nodes; shared nodes are depicted with striped white-and-gray fill. Figure \ref{subfig:BAID1} represents an arbitrary, sequential game. The dashed informational arc between the decision nodes indicates that Player $D$ takes an action after having observed Player $Z$'s choice. Both player's utilities are, in turn, affected by the players' joint action profile and the outcome of the uncertainty $X$. For more information on BAIDs, we refer the interested reader to \citet{koller2003multi}.

Figure \ref{subfig:BAID2} illustrates how an ARA analysis decomposes this competitive interaction. Namely, assuming the ARA is supporting Player $Z$, Player $D$ is treated as, simply put, another source of uncertainty. Whereas a multitude of opponent models may be leveraged within the ARA framework \citep{albrecht2018autonomous}, most research adopts a recursive reasoning approach. This implies that a corresponding ARA reduction is constructed and solved from Player $D$'s perspective; however, due to Player $Z$'s uncertainty about their opponents beliefs, multiple instantiations of this reduction are solved to construct an empirical, probability distribution over Player $D$'s actions. Once this distribution over Player $D$'s actions has been identified, solving Player $Z$'s problem can be accomplished via standard decision-analytic practices. Alternatively, \citet{banks2011adversarial} illustrated how, under low information conditions, zeroth-order ARA allows for the direct elicitation of beliefs on the opponent's behavior.

Due to its generality, ARA can be applied to any competitive setting. A plurality of the ARA literature focuses on its application to physical security settings, e.g., counter-terrorism and military operations. However, recent research highlights its promise for adversarial machine learning as well \cite[e.g., see][]{naveiro2019classification, rios2023adversarial, gonzalez2021hypothesis}. Canonical AML techniques are rooted in game-theoretic analysis, and inherit the associated common knowledge assumption. An ARA approach allows one to loosen this assumption when it is inappropriate, e.g., for the security settings considered herein.

\section{Hidden Markov Model Corruption Problems} \label{secProbs}

We consider herein an attacker, i.e., Player $Z$, attempting to thwart inference conducted on a HMM by a decision maker, i.e., Player $D$. Given that many HMM-inference techniques utilize batch data, i.e., a full observation sequence $\{x_t\}_{t \in \mathcal{T}}$, we assume the attacker seeks to modify this information so that, when utilized by the decision maker, the resulting inference somehow benefits Player $Z$. Whereas the attacker knows the decision maker is utilizing an HMM, they are uncertain of its exact parameterization (i.e., in the terms of \cite{biggio2018wild}, we consider a grey-box setting) and describe their beliefs probabilistically in a Bayesian manner. Moreover, the attacker's attempts at data corruption are subject to error and are not guaranteed to be successful. Corruption attacks are also accompanied by an associated risk of discovery, implying that Player $Z$ must balance the risks and rewards of their attacks. For this initial research, the decision maker is assumed to utilize the standard HMM inference procedures discussed in Section \ref{secBack}; that is, as with many HMM applications currently in use, Player $D$ has not hardened their algorithms against potential attacks.

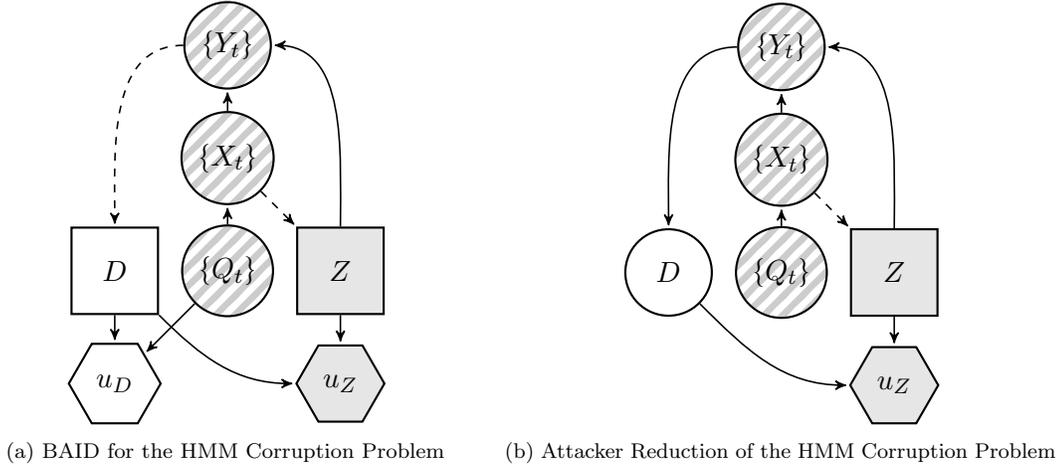
\begin{figure}[htbp]
\begin{subfigure}{.5\textwidth}
\centering
\begin{tikzpicture}[->,>=stealth',shorten >=1pt,auto,node distance=1.5cm,
                    semithick]
  \tikzstyle{uncertain}=[circle,pattern=stripes,
                                    pattern color=gray!40,
                                    thick,
                                    minimum size=1.15 cm,
                                    draw=black]
  \tikzstyle{uncertainX}=[circle,pattern=stripes,
                                    pattern color=gray!40,
                                    thick,
                                    minimum size=1.15 cm,
                                    draw=black]
  \tikzstyle{uncertainY}=[circle,pattern=stripes,
                                    pattern color=gray!40,
                                    thick,
                                    minimum size=1.15 cm,
                                    draw=black]
  \tikzstyle{Attacker_utility}=[regular polygon,regular polygon sides=6,
                                    thick,
                                    minimum size=1.15 cm,
                                    draw=black,
                                    fill=gray!20]
  \tikzstyle{Defender_utility}=[regular polygon,regular polygon sides=6,
                                    thick,
                                    minimum size=1.15 cm,
                                    draw=black,
                                    fill=white]
  \tikzstyle{Attacker_decision}=[rectangle,
                                    thick,
                                    minimum size=1.15  cm,
                                    draw=black,
                                    fill=gray!20]
  \tikzstyle{Defender_decision}=[rectangle,
                                    thick,
                                    minimum size=1.15  cm,
                                    draw=black,
                                    fill=white]
  \tikzstyle{texto}=[label]			
  \node[uncertain](D)   {$\{Q_t\}$};
  \node[Defender_decision] (A) [left of=D]  {$D$};
  \node[Attacker_decision] (B) [right of=D] {$Z$};
  \node[uncertainX] (X) [above of=D] {$\{X_t\}$};
  \node[uncertainY] (Y) [above of=X] {$\{Y_t\}$};
  \node[Defender_utility]  (C) [below of=A] {$u_D$};
  \node[Attacker_utility]  (E) [below of=B] {$u_Z$};
  \path (A) edge    node {} (C)
            edge[out=-45, in = 180]    node {} (E)
        (B) edge    node {} (E)
            edge[out=90, in=0]  (Y)
        (D) edge    node {} (C)
            edge    node {} (X)
        (X) edge    node {} (Y)
            edge[dashed]      node {} (B)
        (Y) edge[out=180, in=90, dashed] node {} (A);
\end{tikzpicture}
\caption{BAID for the HMM Corruption Problem} \label{fig:batchBAID}
\end{subfigure}
\begin{subfigure}{.5\textwidth}
\centering
\begin{tikzpicture}[->,>=stealth',shorten >=1pt,auto,node distance=1.5cm,
                    semithick]
  \tikzstyle{uncertain}=[circle,pattern=stripes,
                                    pattern color=gray!40,
                                    thick,
                                    minimum size=1.15 cm,
                                    draw=black]
  \tikzstyle{Defender_uncertainty}=[circle,                                   fill=white,
                            thick,
                            minimum size=1.15 cm,
                                    draw=black]
  \tikzstyle{Attacker_utility}=[regular polygon,regular polygon sides=6,
                                    thick,
                                    minimum size=1.15 cm,
                                    draw=black,
                                    fill=gray!20]
  \tikzstyle{Defender_utility}=[regular polygon,regular polygon sides=6,
                                    thick,
                                    minimum size=1.15 cm,
                                    draw=black,
                                    fill=white]
  \tikzstyle{Attacker_decision}=[rectangle,
                                    thick,
                                    minimum size=1.15  cm,
                                    draw=black,
                                    fill=gray!20]
  \tikzstyle{Defender_decision}=[rectangle,
                                    thick,
                                    minimum size=1.15  cm,
                                    draw=black,
                                    fill=white]
  \tikzstyle{texto}=[label]			
  \node[uncertain](D)   {$\{Q_t\}$};
  \node[Defender_uncertainty] (A) [left of=D]  {$D$};
  \node[Attacker_decision] (B) [right of=D] {$Z$};
  \node[uncertain] (X) [above of=D] {$\{X_t\}$};
  \node[uncertain] (Y) [above of=X] {$\{Y_t\}$};
  \node[Attacker_utility]  (E) [below of=B] {$u_Z$};
  \path (A) edge[out=-45, in = 180]    node {} (E) 
        (B) edge    node {} (E)
            edge[out=90, in=0]  (Y)
        (D) edge    node {} (X)
        (X) edge    node {} (Y)
            edge[dashed]      node {} (B)
        (Y) edge[out=180, in=90] node {} (A);
\end{tikzpicture}
\caption{Attacker Reduction of the HMM Corruption Problem} \label{fig:batchBAID_reduced}
\end{subfigure}
\caption{HMM Corruption Problem} 
\end{figure}

Figure \ref{fig:batchBAID} provides a BAID which graphically depicts this interaction\footnote{In select problem variants (e.g., distribution disruption), a \textit{functional arc} between $\{X_t\}$ and $u_Z$ may need to be considered. We discuss such problems subsequently.}. Akin to the real-world scenarios discussed by \citet{Krasser2023}, the attacker is assumed to have either infiltrated the decision maker's information system or controls the input data. They desire to use this access, along with intelligence about the form of $\mathcal{Q}$ and $\mathcal{X}$, to manipulate inference while maintaining data plausibility. The series of true latent states, $\{Q_t\}_{t\in\mathcal{T}}$, is unknown to each player; however, it affects the observations, $\{x_t\}_{t\in\mathcal{T}}$. Player $Z$ is able to first view this information and, in accordance with their own self interest, attempt to corrupt it by altering observations. That is, if it will maximize their expected utility, Player $Z$ may attempt to change a set of $x_t$-variables but, because these attacks are subject to error, the attack may not be entirely successful. Via Player $Z$'s interaction with $\{x_t\}_{t\in\mathcal{T}}$, a new, potentially perturbed series of observations, $\{y_t\}_{t\in\mathcal{T}}$ is utilized by the decision maker (i.e., Player $D$) as the basis of their inference. Such attacks are particularly relevant when the environment and the sensors determine the emissions and latent variables (e.g., quality control or fault identification on assembly lines).

Should Player $Z$ have perfect knowledge about the HMM parameterization, then the resulting HMM corruption problem would be deterministic from the attacker's perspective. Herein, we adopt an alternative approach that we contend is more realistic. Namely, an attacker can quite readily ascertain via nefarious means the class of machine learning algorithm utilized in application by a decision maker, but discovering its exact parameterization is a more difficult task. According to such uncertainties, Figure \ref{fig:batchBAID_reduced} presents an ARA reduction of the previously described BAID from the attacker's perspective. Player $Z$'s beliefs about the decision maker's parameterization  (i.e., $a_{i,j}$, $b_{i,k}$, and $\pi_i$) are encapsulated by the uncertainty node on Player $D$'s decision. 

Although Figure \ref{fig:batchBAID_reduced} provides an intuitive graphical means of depicting the attacker's problem, further notation is required to set forth an authoritative formulation. Table \ref{tab:notation} details the additional notation leveraged within this section to do so. Therein, variable indices are represented at their finest level of granularity; however, hereafter, arrays of decision variables are denoted via the absence of a subscript (e.g., $z_t$ contains $z_{t,k}$ over every $k \in \mathcal{X}$). The same convention is used for the probability matrices, e.g., row $i$ of $A$ is denoted as $a_i=(a_{i,1},\ldots, a_{i,|\mathcal{Q}|})$.

\begin{table}[htbp!]
\centering
\caption{Summary of Additional Notation for HMM Corruption Problems} \label{tab:notation}
\resizebox{0.9\textwidth}{!}{
\begin{tabular}{c p{.9\textwidth}}
\textbf{Notation} & \textbf{Definition} \\ \hline
$u_Z(\cdot)$ & Player $Z$'s utility for a given set of decision variables and realized uncertainties\\
$z_{t,k}$ & Binary decision variable equal to 1 if attacker inserts $k$ at $t$, and 0 otherwise \\ 
$w_1$ & Attacker's objective-function weight on decision maker's inference\\ 
$w_2$ & Attacker's objective-function weight on data corruption costs\\ 
$\rho_{t,k}$ & Random variable equaling 1 if insertion of $k$ at $t$ is successful, and 0 otherwise\\
$y_t$ & Perturbed observation at $t$ viewed by decision maker \\
$\alpha_{t,i}$ & Standard forward probabilities calculated with $\{y_t\}_{t\in\mathcal{T}}$\\
$\beta_{t,i}$ & Standard backward probabilities calculated with $\{y_t\}_{t\in\mathcal{T}}$\\
$\delta_{t,i}$ & Standard Viterbi probabilities calculated with $\{y_t\}_{t\in\mathcal{T}}$\\
$P_\rho$ & Joint probability mass function of $\rho$-variables having support $\mathcal{P}$\\
$g_A$ & Joint density over entries in $A$ having support $\mathcal{A}$\\
$g_B$ & Joint density over entries in $B$ having support $\mathcal{B}$\\
$g_\pi$ & Dirichlet density over entries in $\pi$ having support $\Pi$ \\
$g_\omega$ & Joint density over all random variables having support $\Omega$\\
\hline
\end{tabular} }
\end{table}

\subsection{Corrupting Filtering and Smoothing Inference}
The attacker's problem described previously is simultaneously stochastic, combinatorial, and nonlinear. These characteristics are therefore present in Problem \textbf{P1} which represents an attacker with general, multi-objective utility over the corruption of a decision maker's  filtering or smoothing distributions at time $t'$.  When encountered with such a problem, the attacker should solve 

\small
\begin{subequations}
\begin{align}
\allowdisplaybreaks
\textbf{P1}:   &\max_{z}  && u_Z(z,\alpha_{t'},\beta_{t'}) = \mathbbm{E}\left[w_1 f_1(\alpha_{t'},\beta_{t'}) -  w_2 f_2(z) \right] \notag \\
    &\text{s.t.} \  && \sum_{k \in \mathcal{X}} z_{t,k} =1, \ \forall t \in \mathcal{T}, \label{eqOneAttackSelection}\\
   & && \alpha_{1,i} = \pi_i \left( \sum_{k \in \mathcal{X}} z_{1,k} (b_{i,k} \rho_{1,k} +  b_{i,x_1}(1-\rho_{1,k})) \right), \ \forall i \in \mathcal{Q}, \label{eqForward1} \\
   & && \alpha_{t,i} = \sum_{j \in \mathcal{Q}} \alpha_{t-1,j} a_{j,i} \left( \sum_{k \in \mathcal{X}} z_{t,k} (b_{i,k} \rho_{t,k} +  b_{i,x_t}(1-\rho_{t,k})) \right) , \ \forall t \in \mathcal{T} \setminus \{1\}, \ i \in \mathcal{Q}, \label{eqForward2}\\
    & && \beta_{|\mathcal{T}|,i} = 1, \ \forall i \in \mathcal{Q}, \label{eqBackward1} \\
   & && \beta_{t-1,i} = \sum_{j \in \mathcal{Q}} \beta_{t,j} a_{i,j} \left( \sum_{k \in \mathcal{X}} z_{t,k} (b_{j,k} \rho_{t,k} +  b_{j,x_t}(1-\rho_{t,k})) \right) , \ \forall t \in \mathcal{T} \setminus \{1\}, \ i \in \mathcal{Q}, \label{eqBackwardProbRecursion} 
\end{align}
\end{subequations}
\normalsize

\noindent such that

\begin{align*}
    \rho_{t,k} & \sim \mathscr{B}(\lambda_k), \forall t \in \mathcal{T}, \ k \in \mathcal{X}, \\
    a_{i} & \sim \mathscr{D}(\xi_i), \forall i \in \mathcal{Q}, \\
    b_{i} & \sim \mathscr{D}(\zeta_i), \forall i \in \mathcal{Q}, \\
    \pi & \sim \mathscr{D}(\upsilon),
\end{align*}

\noindent where $\mathscr{B}(\cdot)$ and $\mathscr{D}(\cdot)$ are shorthand for the Bernoulli and Dirichlet distributions characterized by the provided parameters. We note that attack success, i.e., $\rho_{t,k}$, is stationary across $t$ but may vary in $k$. Similarly, uncertainty over the transition and emission probabilities may vary in $i$. For notational convenience, the joint probability mass function of all $\rho$-variables is denoted by $P_\rho$ over support $\mathcal{P}$,  the joint density of the transition matrix rows is $g_A$ with support $\mathcal{A}$, the joint density of the emission matrix rows is $g_B$ with support $\mathcal{B}$, and the Dirichlet distribution over $\pi$ is denoted by $g_\pi$ with support $\Pi$. The joint density of all random variables is denoted by $g_\omega$ with support  $\Omega=\mathcal{P} \times \mathcal{A} \times \mathcal{B} \times \Pi$.

The objective function of Problem P1 is the expected value of a linear combination of $f_1(\alpha_{t'},\beta_{t'})$ and $f_2(z)$ whereby the former captures the attacker's utility from the decision maker's inference and the latter represents the utility associated with their corruption decisions. That is, the weighted-sum method from multicriteria optimization is leveraged \citep{ehrgott2005multicriteria}. Constraint \eqref{eqOneAttackSelection} ensures that the attacker only selects a single $k \in \mathcal{X}$ to insert into the perturbed observation vector at each time. Note that, if $x_t=k$ and $z_{t,k}=1$, the attacker is not altering $x_t$ and $y_t=k$ with certainty, i.e., this observation is not being changed. Alternatively, Constraints \eqref{eqForward1}--\eqref{eqBackwardProbRecursion} ensure the correct calculation of both the forward and backward probability recursions. Although visually distinct, these constraints are conceptually similar to the equations provided in Section \ref{secBackHMM}. The summand in \eqref{eqForward1}--\eqref{eqBackwardProbRecursion} over $k \in \mathcal{X}$, ensures that the forward (backward) probabilities are calculated correctly with the realized $\{y_t\}_{t \in \mathcal{T}}$. Moreover, provided a realized set of attack-success outcomes (i.e., $\rho_{t,k}$), this sequence of perturbed observations can readily be calculated via

$$ y_{t} = \sum_{k' \in \mathcal{X}} z_{t,k'}  \left(\rho_{t,k'} k'  + 
(1-\rho_{t,k'})x_{t} \right), \ \forall t \in \mathcal{T}. $$

From inspecting Problem P1, its difficulty becomes immediately apparent. By expanding the recursions in Constraint \eqref{eqForward2} and \eqref{eqBackwardProbRecursion}, it can be observed that the resulting functions are nonlinear for any non-trivial $\mathcal{T}$; both the forward and backward probabilities expand to, potentially, high-order polynomial functions. Adding further complexity is the fact that all of the parameters in Constraints \eqref{eqForward1}--\eqref{eqBackwardProbRecursion} are unknown and subject to probabilistic uncertainty. Furthermore, whereas Constraint \eqref{eqOneAttackSelection} is deterministic and linear, it highlights the combinatorial nature of the attacker's problem. 
This complexity may be further exacerbated by the specific functional form of the attacker's multi-objective utility function. 

Herein, we explore a subset of the most interesting utility functions.  Multitudinous options exist to parameterize the component utility function $f_2(z)$. One straightforward and flexible option is

\begin{equation}
    f_2(z) = \sum_{t \in \mathcal{T}}\sum_{k \in \mathcal{X}: x_t \ne k} z_{t,k}.
\end{equation}

\noindent This simple alternative penalizes any corruption conducted by the attacker, and ensures this risk is appropriately represented in the attacker's utility function by its product with $w_2$. Higher-order polynomials in the $z_{i,k}$-variables could also be leveraged if the Player $Z$'s attack costs are not constant. A similar diversity of functional representation is associated with $f_1(\alpha_{t'},\beta_{t'})$. Subsequent sections explore this dynamic in greater detail whereby we provide alternative structures according to varying attacker intentions. 

Given that Problem P1 contains unknown parameters described probabilistically, it is clearly a stochastic programming problem. The recursions within Constraint \eqref{eqForward1} -- \eqref{eqBackwardProbRecursion} are defined to ensure clarity of communication and highlight the problem's relationship to canonical HMM algorithms but are not strictly necessary. That is, the recursion may be written explicitly in the objective function.

\subsubsection{State-Attraction and State-Repulsion Problems}

In some settings, the attacker may only be interested in the decision maker's conditional distribution over a latent state at a particular time $t'$. For example, if the latent state represented the attacker's position and they wish to evade detection by the decision maker, it is in the attacker's best interest to minimize the decision maker's belief about this true position. Similarly, it is conceivable that an attacker may wish to lure the decision maker into believing a specific latent state $i'$ occurred at time $t'$. In either setting, these objectives can be modeled using

\begin{align*}
    f_1(\alpha_{t'}, \beta_{t'}) = 
    c\left(\frac{\alpha_{t',i'}\beta_{t',i'}}{\sum_{j \in \mathcal{Q}} \alpha_{t',j} \beta_{t',j} } \right)
\end{align*}

\noindent where $c=1$ in a state-attraction problem and $c=-1$ in a state-repulsion problem. Therefore, the attacker will maximize or minimize, respectively, the 
smoothing probability provided in Equation \eqref{eqSmooth}.

This component utility function is associated with an attacker who is concerned with a single latent state. While useful, this may not always properly characterize the attacker's intent. Some attackers may wish to alter the decision maker's inference over multiple latent states, e.g., potentially the entirety of $\mathcal{Q}$ as presented subsequently.

\subsubsection{Distribution-Disruption Problems}

The attacker may be interested in disrupting the decision maker's beliefs at time $t'$ across $\mathcal{Q}$ to the maximum extent possible. More specifically, the attacker may wish to maximize the distance between the filtering (or smoothing) distributions under $\{x_t\}_{t\in\mathcal{T}}$ and $\{y_t\}_{t\in\mathcal{T}}$. Numerous options exist to characterize the distance between the uncorrupted and corrupted distributions, e.g., the Kullback-Leibler divergence and the Hellinger distance  \citep{cha2007comprehensive}. Generally speaking, a distribution-disruption problem can be formulated by setting $f_1(\cdot)$ equal to the desired distance measure.

Let $\hat{\gamma}_{t,i}$ represent probability of state $i$ at some time $t$ given the uncorrupted data, $\{x_t\}_{t\in\mathcal{T}}$.  Using this notation, the Kullback-Leibler divergence between the distributions induced by $\{x_t\}_{t\in\mathcal{T}}$ and $\{y_t\}_{t\in\mathcal{T}}$ is

\begin{align*}
    f_1(\alpha_{t'}, \beta_{t'}) = \sum_{i \in \mathcal{Q}} \hat{\gamma}_{t',i} \left(\log (\hat{\gamma}_{t',i}) - \log\left(  \frac{\alpha_{t',i}\beta_{t',i}}{\sum_{j \in \mathcal{Q}} \alpha_{t',j} \beta_{t',j} } \right) \right).
\end{align*}

\noindent The above can be used to identify a maximally perturbed distribution akin to an ill-performing $M$-projection; however, the order of the distributions may be reversed in the Kullback-Leibler divergence to identify a modified $I$-projection \citep{koller2009probabilistic} as well. Alternatively, if the Hellinger distance is utilized, then one may set

\begin{align*}
    f_1(\alpha_{t'}, \beta_{t'}) = \frac{1}{\sqrt{2}} \sqrt{\sum_{i \in \mathcal{Q}} \left( (\hat{\gamma}_{t',i})^{\nicefrac{1}{2}} - \left(  \frac{\alpha_{t',i}\beta_{t',i}}{\sum_{j \in \mathcal{Q}} \alpha_{t',j} \beta_{t',j} } \right)^{\nicefrac{1}{2}} \right)^2}.
\end{align*}

\noindent Other statistical distances calculated via an optimization step may also be used (e.g., the Wasserstein distance or the Kolmogorov–Smirnov statistic); however, when adopting such an approach, Problem P1 may need to be augmented with the requisite, distance-specific constraints. Nevertheless, we henceforth utilize the Kullback-Leibler divergence in our exploration, but such a selection is merely a matter of attacker preference.

\subsection{Corrupting Decoding Predictions}

The probabilistic dynamics of the underlying HMM suggests that, when corrupting decoding predictions, the attacker is once more confronted with a non-linear, combinatorial, stochastic problem. Problem \textbf{P2} presents a mathematical programming model for an attacker trying to disrupt a decision maker inferring the most likely sequence of latent states given $\{y_t\}_{t\in\mathcal{T}}$.

\small
\begin{subequations}
\begin{align}
\allowdisplaybreaks
\textbf{P2}:   &\max_{z}  && u_Z(z,\delta) = \mathbbm{E}\left[w_1 f_1(\delta) -  w_2 f_2(z) \right]\\
 &\text{s.t.} \  && \sum_{k \in \mathcal{X}} z_{t,k} =1, \ \forall t \in \mathcal{T}, \\
   & && \delta_{1,i} = \pi_i \left( \sum_{k \in \mathcal{X}} z_{1,k} (b_{i,k} \rho_{1,k} +  b_{i,x_1}(1-\rho_{1,k})) \right), \ \forall i \in \mathcal{Q}, \\
   & && \delta_{t,i} = \max_{j \in \mathcal{Q}} \delta_{t-1,j} a_{j,i} \left( \sum_{k \in \mathcal{X}} z_{t,k} (b_{i,k} \rho_{t,k} +  b_{i,x_t}(1-\rho_{t,k})) \right) , \ \forall t \in \mathcal{T} \setminus \{1\}, \ i \in \mathcal{Q}. \label{eqViterbiMax}
\end{align}
\end{subequations}
\normalsize

\noindent where

\begin{align*}
    \rho_{t,k} & \sim \mathscr{B}(\lambda_k), \forall t \in \mathcal{T}, \ k \in \mathcal{X}, \\
    a_{i} & \sim \mathscr{D}(\xi_i), \forall i \in \mathcal{Q}, \\
    b_{i} & \sim \mathscr{D}(\zeta_i), \forall i \in \mathcal{Q}, \\
    \pi & \sim \mathscr{D}(\upsilon).
\end{align*}

As with Problem P1, Problem P2's objective function is the expected value of  a linear combination of two component utility functions. The cost of an attack vector, $f_2(z)$, may be of a similar form as those discussed for Problem P1. However, since the decision maker is assumed to be solving a decoding problem, the attacker's component utility on their inference, $f_1(\delta)$, differs from Problem P1. The storage of forward and backward probabilities is no longer strictly required, and the Viterbi probabilities, $\delta_{t,i}$, are computed instead. However, in so doing, the constraints of Problem P2 are further complicated via the inclusion of a maximum operator, i.e., Constraint \eqref{eqViterbiMax}. 

Given a realized set of uncertainties (i.e., $\rho$, $A$, $B$, $\pi$), the most-likely state sequence from the decision maker's perspective can be calculated via backtracking. However, for notational simplicity, the requisite back pointers that store the states maximizing Constraint \eqref{eqViterbiMax} are excluded from Problem P2; they are not necessary to identify the attacker's optimal solution for the following problems.

\subsubsection{Path-Attraction and Path-Repulsion Problems}

As with Problem P1, there exist multiple component utility functions that may be used for $f_1(\delta)$. However, we are primarily concerned herein with path-attraction and path-repulsion problems. In a path-attraction problem, the attacker wishes to encourage the decision maker to believe in a particular sequence of latent states. This is accomplished by setting

\begin{align*}
        f_1(\delta) = \sum_{t \in \mathcal{T}}\sum_{i \in \mathcal{Q}} c_{t,i}\delta_{t,i}
\end{align*}

\noindent such that $c_{t,i}>0$ for states the attacker would like to encourage at each $t \in \mathcal{T}$. Alternatively, a path-repulsion problem is characterized by $c_{t,i}<0$. The attacker may identify these sequences of latent states exogenously or base them off their expectation of $\{q_t\}_{t\in \mathcal{T}}$ under the true data $\{x_t\}_{t\in \mathcal{T}}$. Likewise, it is straightforward to combine these two frameworks to simultaneously attract and repulse a subset of latent state sequences. By their nature, the $\delta_{t,i}$-variables decrease in $t$, implying that, unless offset by increasing $c_{t,i}$-values, latent states visited later in the sequence would become relatively less important. If such an effect is not desirable, an effective method is to set $c_{t,i}=(\sum_{j \in \mathcal{Q}} \delta_{t,j})^{-1}$ for any state $i$ the attacker wishes to encourage and set $c_{t,i}=-(\sum_{j \in \mathcal{Q}} \delta_{t,j})^{-1}$ for any state $i$ the attacker wishes to discourage. Notably, by working as such with the $\delta_{t,i}$-probabilities, the attacker does not necessarily induce the Viterbi algorithm to output the desired path; however, it is an approximation that is clearly associated with such induction.

\section{Solution Methodologies} \label{SecSolution}

The HMM structure utilized by the decision maker presents multiple, simultaneous difficulties. In the filtering, smoothing, and decoding settings, the attacker is presented with a non-linear objective function. Moreover, this difficulty is compounded by a set of decision variables of combinatorial cardinality, as well as a lack of knowledge that induces a stochastic programming problem. Of the commercial optimization solvers available, only global solvers are capable of accepting such problems as input; however, as shown by \citet{caballero2018challenges}, even these methods are not foolproof in complex situations such as the one described in this paper. Finally, as a stochastic programming problem, solving Problems P1 and P2 coincides with maximizing an expectation, but its non-linearity limits the usefulness of canonical methods (e.g., sample-average approximation). 

It is therefore apparent that customized solution techniques are required for Problems P1 and P2. Given the success of standard HMM algorithms, their modification to our perturbed setting would be ideal. Unfortunately, the relative efficiencies of the forward, backward, and Viterbi algorithms draw from their dynamic-programming structure, and this structure is dependent upon having observed a sequence of observations (e.g., $\{x_t\}_{t \in \mathcal{T}}$) upon which to base calculations. While desirable, it is unclear how (or if), the dynamic-programming structure of the aforementioned algorithms can be adapted to simultaneously optimize the attacker's actions and calculate the HMM's joint probabilities. The tractable identification of an optimal attack vector is, at this juncture, a dubious prospect; therefore, we set forth alternative means the attacker may leverage to maximize their expected utility based on their subjective beliefs.

Within this section, we provide customized algorithms for solving variants of Problems P1 and P2, with minor modifications. The first solution method is a heuristic inspired by the \textit{ranking-and-selection} problem \citep{powell2019unified}. The second is an augmented-probability-simulation technique rooted in inhomogeneous Markov chain Monte Carlo sampling. The third is a Monte-Carlo enumeration technique having complete and random-greedy variants. When allocated enough computational resources, the complete variant will converge to the optimal solution of Problems P1 or P2. Based on this property, this algorithm serves as the benchmark solution method. Notably, these attacks may be used for each of the problems from Section \ref{secProbs} with minor modifications. Likewise, they are applicable to alternative prior distributions to those assumed in the previous section; an analytic prior is not even necessary, as long as the attacker can sample from the associated distribution.

\subsection{Ranking-and-Selection Heuristic} \label{secR&S}

Once the attacker has probabilistically codified their beliefs about the decision maker's problem, Problems P1 and P2 can be recast via the universal canonical model set forth by \citet{powell2019unified}. More specifically, the HMM corruption problem is conceptualized as a variant of the ranking-and-selection problem whereby the attacker is asked to estimate the optimal $z^* \in \mathcal{Z}=\{z: z_{t,j} \in \{0,1\}, \forall (t,j) \in \mathcal{T}\times \mathcal{X}; \sum_{j \in \mathcal{Q}} z_{t,j}=1, \forall t \in \mathcal{T} \}$ after having run $N$ experiments according to some policy $\eta$. These experiments are simulations of potential realities based upon the attacker's beliefs about attack success and the HMM parameterization. Traditional ranking-and-selection (R\&S) problems consider a finite action space of relatively small cardinality, thereby enabling estimates of this expectation to be stored tabularly after each experiment $n$. In juxtaposition, the action-space cardinality may be exceedingly large in our HMM corruption problems. Therefore, in what follows, we utilize a function-approximation approach to estimate an action's expected value. This method is highly flexible, allowing for use of various combinations of function-approximation methods and combinatorial-optimization routines.

Powell's universal canonical model requires the following elements be defined. Sequences of experimental values and estimates are denoted by $n$ superscripts to distinguish them from the temporal, HMM sequences denoted by $t$ subscripts. 

\begin{itemize}
    \item \textit{State Variables}: The system state prior to conducting experiment $n$ is denoted by $\theta^n$, and determines the attacker's beliefs about each action's expected value.  These beliefs are codified via a functional approximation $\hat{\mu}(z^n|\theta^n)$, i.e., $\theta^n$ are the parameters of the attacker's functional approximation prior to experiment $n$.  The exact structure of $\theta^n$ depends upon the approximation utilized; however, under any setting, the set of all possible state variables is denoted by $\Theta$.
    \item \textit{Decision Variables}: For each experiment $n=1,...,N$, the attacker selects $z^n \in \mathcal{Z}$.
    \item \textit{Exogenous Information}: Given some attack vector $z^n$, the attacker's payoff is determined by the decision maker's inference on $\{y_t\}_{t\in \mathcal{T}}$ which is in turn affected by the realized exogenous information, i.e., the random $\rho$-, $A$-, $B$-, and $\pi$-variables. An instance of this exogenous information during experiment $n$ is referred to as $\omega^n \in \Omega$.
    \item \textit{Objective Function}: For Problems P1 and P2, the attacker's objective is to maximize their expected reward by selecting an experimental policy $\eta$ that respectively solves
    \begin{equation*}
        \max_{\eta} \mathbbm{E}\left[u(Z^\eta(\theta^{N+1}),\alpha_{t'},\beta_{t'}) \right]
    \end{equation*}
    and 
    \begin{equation*}
        \max_{\eta} \mathbbm{E}\left[u(Z^\eta(\theta^{N+1}),\delta) \right]
    \end{equation*}
    
    \item \textit{System Model}: Given that the system state, $\theta^n$, denotes the parameters underpinning $\hat{\mu}(z^n|\theta^n)$, the system model, $S^M(\theta^n, z^n, \omega^n)$, corresponds to the sequential update of $\theta^n$ after each experiment, i.e., $S^M: \Theta  \times \mathcal{Z} \times \Omega \to \Theta$. The system model reflects the sequential optimization of the $\theta^n$ parameters to $\theta^{n+1}$ after each experiment $n$. Example system models $\theta^n$-updates include stochastic gradient descent and recursive least squares, among others.
\end{itemize}

Given this formulation, Algorithm \ref{algRandS} provides an overarching 
R\&S heuristic that is applicable to either problem under any of the attacker utility functions presented in Section \ref{secProbs}. The attacker must input their beliefs about the associated uncertainities (i.e., $g_\omega)$, their initial functional-approximation parameterization (i.e., $\theta^1)$, and the number of experiments they wish to conduct (i.e., $N$). The R\&S heuristic concludes by outputting a recommended attack, $\hat{z}^*$. Note that Step \ref{RandS_stepX} may be omitted for most problems; it is only required in select settings, e.g., distribution-disruption problems.

\small
\begin{algorithm}[!htbp]
\caption{Ranking-and-Selection Heuristic} 
\label{algRandS}
\begin{algorithmic}[1]
\State{\textbf{Input}: $g_\omega$, $\theta^1$, $N$}
\State{\textbf{Output}: $\hat{z}^*$ }
    \For{$ n= 1, \ldots, N$}  \Comment{May alternatively use clock-time limit}
        \State{Select $z^n=Z^{\eta}(\theta^n)$}  \label{stepIDaction}
        \State{Sample $\omega^n \sim g_\omega$}
\State{Solve decision maker's HMM problem with $\{x_t\}_{t\in\mathcal{T}}$} \Comment{If required} \label{RandS_stepX}
        \State{Use $\omega^n$ and $z^n$ to determine $\{y_t\}_{t\in\mathcal{T}}$} \label{RandSOpt}
        \State{Solve decision maker's HMM problem with $\{y_t\}_{t\in\mathcal{T}}$ and $\omega^n$, and compute $u_Z(\cdot)$ }
        \State{Update $\theta^{n+1} = S^M(\theta^n, z^n, \omega^n)$}
    \EndFor
    \State{Select $\hat{z}^*=Z^{\eta}(\theta^{N+1})$} \label{stepIDaction2}
  \State \Return $\hat{z}^*$ 
\end{algorithmic}
\end{algorithm}
\normalsize

The flexibility of the formulation allows for the construction of multitudinous R\&S-heuristic variants.
The policy $\eta$ may adopt multiple structural forms \citep[i.e., see][]{powell2019unified}. Greedy or $\varepsilon$-greedy policies on $\hat{\mu}(z^n|\theta^n)$ are sensible alternatives prevalent in the R\&S literature; however, since $|\mathcal{Z}|$ can inhibit the use of complete enumeration techniques, it may be necessary to identify the greedy policy via integer programming or analogous heuristic methods (e.g., genetic algorithms). Moreover, as with canonical reinforcement learning methods, the function approximation, $\hat{\mu}(z^n|\theta^n)$, may be based upon any statistical, regression model (e.g., regression trees, neural networks, etc.) with a correspondingly tailored system model. For example, if $\hat{\mu}(z^n|\theta^n)$ is a linear regression model then, akin to \citet{powell2007approximate}, it is sensible to utilize recursive least squares as a system model; however, if $\hat{\mu}(z^n|\theta^n)$ is a neural network then system model updates via stochastic gradient descent are more applicable\footnote{In Section \ref{secTRA}, we illustrate two R\&S variants based on neural networks, that differ in identification of the greedy action, i.e. Monte Carlo tree search and simulated annealing.}. Ultimately, the attacker is likely best served by tuning these qualitative hyperparameters via some sort of experimental design, e.g., see \citet{jenkins2021approximate}.

\subsection{Augmented Probability Simulation}
The problems described in Section \ref{secProbs} can each be recast as indentifying a solution to

\begin{eqnarray} \label{eq:general_optim}
\max_z \sum_{\rho} \int_{\mathcal{A}} \int_{\mathcal{B}} \int_{\Pi} u_Z(z, \phi)  g_A(A)g_B(B) g_\pi(\pi) P_{\rho}(\rho) \dd \pi \dd B \dd A 
\end{eqnarray}

\noindent where $\phi=\Phi(z, A, B,  \pi,\rho)$ is determined by $z$, $A,$ $B,$ $\pi$, and $\rho$, and represents the relevant set of parameters 
and decision variables 
for the associated problem (e.g., $\phi = (\alpha_{t'}, \beta_{t'})$ for Problem P1).

One possible means to solve this problem consists of leveraging simulation-based techniques, such as the augmented probability simulation (APS) methods of \citet{bielza1999decision}. This requires defining an augmented distribution on the product space of attacks and uncertainties of the form

\begin{equation} \breve{g} (z, A ,B, \pi, \rho ) \propto u_Z(z, \phi) g_A(A) g_B(B) g_\pi(\pi) P_{\rho}(\rho). \label{eqAPS1}
\end{equation}
\noindent This distribution is well-defined provided that the utility is positive and bounded for all $z$ and $\phi$, which can be achieved by scaling the utility function appropriately.

It is straightforward to see that the global mode of the marginal of $\breve{g} (z, A ,B,  \pi, \rho)$  in $z$, coincides with the solution of Equation \eqref{eq:general_optim}. Therefore, if we sample $(z, A, B,  \pi, \rho ) \sim \breve{g}(z, A ,B, \pi,\rho)$, the sample mode of $z$ approximates the optimal solution.
However, as acknowledged by \cite{muller2004optimal}, identifying the mode of the $z$ samples may be challenging, especially when $z$ is high dimensional. This difficulty is most apparent when the marginal augmented distribution is characterized by flat regions around its global mode. Both of these conditions are likely to emerge in our setting, suggesting an alternative to standard APS methods may be required. 

Following procedures set forth by \cite{muller2004optimal}, mode identification can be facilitated by defining an alternative augmented distribution.  
More specifically, an augmented distribution must be defined such that its marginal in $z$ is proportional to $\left[ \sum_{\rho} \int_{\mathcal{A}} \int_{\mathcal{B}} \int_{\Pi} u_Z(z, \phi) g_A(A) g_B(B) g_\pi(\pi) P_{\rho}(\rho) \dd \pi \dd B \dd A  \right]^H$ where $H \geq 1$ is called the augmentation parameter. Since the associated, marginal-augmented distribution is more peaked, its mode is more easily identified. 
Therefore, we define an augmented distribution by creating $H$ copies of our unknown parameters. This set of copies is denoted by $\{A^h, B^h, \pi^h,\rho^h\}_{h\in \mathcal{H}}$ such that $\mathcal{H}=\{1,2,...,H\}$. The augmented distribution is given by
\begin{equation}
    \breve{g}_H(z, \{A^h, B^h, \pi^h, \rho^h\}_{h\in\mathcal{H}}) \propto \prod_{h\in\mathcal{H}} u_Z\big(z, \phi^h \big) g_A(A^h) g_B(B^h) g_\pi(\pi^h)  P_{\rho}(\rho^h). \label{eqAPS2}
\end{equation}
\noindent where $\phi^h=\Phi(z, A^h, B^h, \pi^h,\rho^h)$. It is straightforward to prove that the marginal of $\breve{g}_H(z, \{A^h, B^h, \pi^h, \rho^h\}_{h\in\mathcal{H}})$ in $z$ is proportional to  $\big[ \sum_{\rho} \int_{\mathcal{A}} \int_{\mathcal{B}} \int_{\Pi} u_Z(z, \phi) g_A(A) g_B(B) g_\pi(\pi)$ $P_{\rho}(\rho) \dd \pi \dd B \dd A  \big]^H$.

Direct sampling from the distribution in Equation \eqref{eqAPS2} is not feasible, since the exact specification of $\breve{g}_H$ is costly. Instead, samples can be obtained using Markov-chain-Monte-Carlo (MCMC) methods \citep{tierney1994markov}. In particular, we design a Metropolis-within-Gibbs sampling approach for use herein. 

Implementing the Gibbs step requires generating samples 
from the full conditionals $\breve{g}_H(z_t|z_{-t} \{A, B, \pi, \rho\}_{h\in\mathcal{H}})$ where $z_{-t}=\{z_{t'}\}_{t'\in\mathcal{T}\setminus \{t\}}$, and to sample from the full conditionals for $A^h$, $B^h$, $\pi^h$ and $\rho^h$ for $h = 1, \dots, H$.  For communicative clarity, let us rewrite Equation \eqref{eqAPS2} as 
\begin{align*}
    \breve{g}_H(z, \{A^{h}, B^{h},  \pi^{h}, \rho^{h}\}_{h\in\mathcal{H}}) &\propto \exp \Bigg \lbrace \sum_{h\in\mathcal{H}} \log[u_Z\big(z, \phi^h \big)] + \log  [ g_A(A^h)] + \log  [ g_B(B^h)] \\
     & \qquad \qquad \qquad  + \log [g_\pi(\pi^h)] + \log [ P_{\rho}(\rho^h)] \Bigg \rbrace.
\end{align*}

\noindent In so doing, it can be readily observed that
\begin{eqnarray*}
    \breve{g}_H(z_t \vert z_{-t}, \{A^h, B^h, \pi^h, \rho^h\}_{h \in \mathcal{H}}) \propto \exp \left( \sum_{h \in \mathcal{H}} \log \left[ u_Z \left( z_t \cup z_{-t}, \phi^h \right) \right]   \right).
\end{eqnarray*}

\noindent This is simply the \textit{softmax} distribution over $\sum_{h \in \mathcal{H}} \log[u_Z\big( z_t \cup z_{-t}, \phi^h \big)]$ for every possible value of $z_t$. Therefore, given $ z_{-t}$ and  $\{A^h, B^h, \pi^h, \rho^h\}_{h \in \mathcal{H}}$,  sampling from the full conditional of $z_t$ is straightforward. 

Conversely, to sample from the full conditionals for $A^h$, $B^h$, $\pi^h$ and $\rho^h$ for $h = 1, \dots, H$, we need to utilize the Metropolis algorithm. For example, noticing that the full conditional for $A^{h}$ is
\begin{align*}
    \breve{g}_H \left( A^{h} \vert z,  \{A^{h'}\}_{h' \in \mathcal{H}\setminus\{h\}}, \{ B^{h'}, \pi^{h'}, \rho^{h'}\}_{h' \in \mathcal{H}} ] \right) \propto \exp  \left(  \log[g_A(A^{h'})] + \log \left[u_Z (z, \phi^h ) \right] \right),
\end{align*}
samples of $A^{h'}$ can be obtained via a Metropolis approach as follows. Assume the current state of the Markov chain is $z, A^h, B^h, \pi^h, \rho^h$ and $\phi^h = \Phi \left( z, A^h, B^h, \pi^h, \rho^h \right)$, then
\begin{enumerate}
    \item Sample $\tilde{A}^h \sim g_A$.
    \item Compute $ \tilde{\phi}^h = \Phi \left( z, \tilde{A}^h, B^h, \pi^h, \rho^h \right)$
    \item Accept $\tilde{A}^h$ with probability
    \begin{equation*}
        \min \left \lbrace 1, \frac{u_Z(z, \tilde{\phi^h})}{u_Z(z, \phi^h)}\right \rbrace.
    \end{equation*}
\end{enumerate}
Samples from the conditionals in $B^h, \pi^h, \rho^h$ can be obtained sequentially in a similar way. Moreover, samples from the full conditionals for $A^h, B^h, \pi^h, \rho^h$ having different $h$-values can also be obtained in parallel. Sampling sequentially from these conditional distributions will asymptotically produce samples from $\breve{g}_H(z, \{A, B, \pi, \rho\}_{h\in \mathcal{H}})$.

However, as suggested by \cite{muller2004optimal} this sampling framework may result in the algorithm getting stuck in local modes, thereby inhibiting the identification of the global mode. This issue can be mitigated by combining this sampling scheme with an annealing schedule that iteratively increases $H$. 
This produces an inhomogeneous Markov chain whose limiting distribution is, under certain conditions \citep{muller2004optimal}, uniform over the set of optimal attacks.  
Algorithm \ref{alg:APS} outlines our APS approximation method which utilizes the augmented distribution $\breve{g}_H$, the aforementioned Metropolis-within-Gibbs sampling approach, and an annealing schedule $\{\mathcal{H}_n\}_{n=1}^{\infty}$ of variable length; note that $e_k$ denotes the standard basis vector in $\mathbbm{R}^{|\mathcal{X}|}$ having a one in the position $k$ and zero elsewhere. At each iteration $n$, a complete vector $z^n$ is stored. The optimal attack vector $\hat{z}^*$ is estimated as the mode of these samples, excluding the burn-in period $n=1,...,n'-1$. Since $z$ is a discrete random variable, it is often useful to approximate this mode for large $\mathcal{T}$, e.g., by estimating the mode of each $z_t$ independently, to avoid excessively large $N$-values. 

To guarantee convergence to the optimal attack(s), the inhomogeneous MCMC simulation needs to be designed so that the stationary distribution of the Markov chain for a fixed $H$ is precisely $\breve{g}_H$. Following well-established literature of MCMC \citep{tierney1994markov}, it is straightforward to see that the limiting distribution of the Metropolis-within-Gibbs Markov chain defined in Algorithm \ref{alg:APS} for each $h \in H_n$ is precisely $\breve{g}_h$. For further information regarding APS, we refer the interested reader to \cite{muller2004optimal} for theoretical proofs of the algorithm's convergence, and to \citet{ekin2014augmented} and \citet{ekin2022augmented} for  
augmented probability simulation applications in decision-theoretic settings.

\small
\begin{algorithm}[!htbp]
\caption{APS Approximation} 
\label{alg:APS}
\begin{algorithmic}[1]
\State{\textbf{Input}: $g_A$, $g_B$, $g_\pi$, $P_{\rho}$ and $\{\mathcal{H}_n\}_{n=1}^{\infty}$, $N$}
\State{\textbf{Output}: $\hat{z}^*$ }
\State Initialize: $z_t$ for $t = 1, \ldots, \mathcal{T}$
\State Set $n = 1$
\While{$n <N$} \Comment{May alternatively use clock-time limit}
\State{Sample $A^h$, $B^h$, $\pi^h$, $\rho^h$ using the Metropolis step $ \forall h \in \mathcal{H}_n$}
    \For{$ t = 1, 2, \ldots, \mathcal{T} $} \
    \State{For $h\in H_n$ and $k \in \mathcal{X}$, determine $\phi^{h} := \Phi \left( e_k \cup z_{-t}), A^h, B^h, \pi^h, \rho^h \right)$}
    \State {Sample $\tilde{z}_t$ from \textit{softmax} distribution described in the Gibbs step}
    \State Update $z_t = \tilde{z}_t$.
    \EndFor
 \State{Set $z^n= \{z_t\}_{t \in \mathcal{T}}$}   
\State Set $n = n+1$

\EndWhile
\State{Estimate $\hat{z}^*$ as the sample mode of $\{z^n:n\ge n'\}$} 
\State \Return $\hat{z}^*$
    
\end{algorithmic}
\end{algorithm}
\normalsize

\subsection{Monte-Carlo Enumeration}

Monte-Carlo sampling techniques are popular solution approaches for standard ARA problems; they serve to numerically approximate otherwise intractable expectation functions. Akin to the proposed R\&S heuristic, our complete Monte-Carlo enumeration (CME) method leverages samples from $g_\omega$ to inform better estimates of the utility of taking each $z \in \mathcal{Z}$; however, the sampling methods utilized are distinct. Both solution approaches randomly sample the attack effects and the decision maker's HMM parameterization (i.e., $\omega^n \sim g_\omega$), but they differ in their evaluation of corruption attacks. Whereas the R\&S heuristic selects a single $z \in \mathcal{Z}$ to evaluate on $\omega^n$ according to $\eta$, the CME technique evaluates every $z \in \mathcal{Z}$ on each $\omega^n$. In so doing, the latter method is able to garner more information from each $\omega^n$, but at the expense of greater computational effort. Psuedocode for this solution method is provided in Algorithm \ref{algMCEnumeration}. The algorithm is applicable to any of the problems presented in Section \ref{secProbs}, but Step \ref{stepID_unpoisoned} is only strictly required in a distribution-disruption problem. This approach serves as a baseline for the other algorithms discussed herein.

To ensure termination on sufficiently large instances, Algorithm \ref{algMCEnumeration} can be modified to form a random-greedy variant.  random-greedy Monte-Carlo enumeration (RME) approach selects an incumbent attack at random and calculates its expected utility. The expected utility of another random attack is then calculated and, if improved, this new attack becomes the incumbent solution. The algorithm terminates when no improvement is identified. While CME may be effective for solving small-size instances, RME is more scalable to larger instances. 

\small 
\begin{algorithm}[!htbp]
\caption{Complete Monte-Carlo Enumeration} 
\label{algMCEnumeration}
\begin{algorithmic}[1]
\State{\textbf{Input}: $g_\omega$, $N$}
\State{\textbf{Output}: $\hat{z}^*$ }
    \For{$ n= 1, \ldots, N$} 
        \State{Sample $\omega^n \sim g_\omega$}
        \State{Solve decision maker's HMM problem with $\{x_t\}_{t\in\mathcal{T}}$} \Comment{If required} \label{stepID_unpoisoned}
        \For{$ z \in \mathcal{Z}$} 
            \State{Use $\omega^n$ and $z$ to determine $\{y_t\}_{t\in\mathcal{T}}$}
             \State{Solve decision maker's HMM problem with $\{y_t\}_{t\in\mathcal{T}}$}
             \State{Set $\hat{u}^{n,z} = u_Z^n(\cdot)$}
        \EndFor
    \EndFor
    \State{ Set $\bar{u}^z$ to the sample average of $\hat{u}^{n,z}$, $\forall z \in \mathcal{Z}$ }
    \State{Identify $\hat{z}^* = \argmax_z \bar{u}^z$}
\end{algorithmic}
\end{algorithm}
\normalsize

Finally, we note that the CME approach is a simpler alternative to the APS scheme. However, several issues are anticipated to arise for larger-sized instances. The most significant concern arises when the maximization of the expected utility, approximated through Monte Carlo sampling, becomes challenging due to the flatness of the approximation with respect to $z$. In such cases, numerical errors inherent in the Monte Carlo approximation of the expected utility may overshadow a relatively small difference in expected utility between the optimal and inferior attacks. In juxtaposition, the APS scheme transforms the simulation-optimization problem into a grand simulation problem using a series of augmented probability models that become more peaked around the optimal attack.

\section{Testing, Results, and Analysis} \label{secTRA}
This section analyzes the effects of HMM corruption on a decision maker's inference and examines the efficacy of the developed heuristics to conduct such attacks. This is accomplished via four blocks of experimentation. Our focus in Section \ref{secIllustrative} is to determine the degree to which data corruption can negatively affect a decision maker's inference. Therein, we explore the effect of varying objective-function weights on the attacker's behavior and illustrate the devastating effects that relatively low-perturbation data-corruption attacks may generate. Analysis within this section also reveals that, although the Monte Carlo enumeration algorithm can identify high-quality solutions, it is computationally infeasible for larger-sized HMMs. Therefore, in Sections \ref{secStrucTest} and \ref{secUncTest} we provide detailed analyses of the use of our R\&S heuristic and the APS approximation routine. Section \ref{secStrucTest} analyzes the effect of the HMM structure on each algorithm's performance, whereas Section \ref{secUncTest} focuses on the impact of uncertainty on solution quality. Section \ref{sec:CaseStudy} provides a case study illustrating the practical relevance of our attacks by attacking an HMM used for part-of-speech tagging.

While the case study leverages a standard laptop for computation to demonstrate real world application feasibility, the remainder of testing was performed using relatively powerful machines. This enabled parallelization and multiple runs of each attack for improved statistical analysis. Such testing was performed with the LOVELACE High Performance Computing (HPC) infrastructure housed at the Institute of Mathematical Sciences of the Spanish National Research Council (ICMAT-CSIC). In particular, 12 nodes were leveraged for experimentation, each equipped with 187GB of RAM and two 32-core, 2.30GHz Intel(R) Xeon(R) processors. 
For additional information regarding the HPC cluster layout and the computational resources available therein, we refer the interested reader to \citet{ICMAT2022}.

\subsection{Efficacy of HMM Corruption}\label{secIllustrative}
The purpose of this section is to illustrate the adverse effects of data corruption on HMM inference and showcase how this relates to the attacker's objectives and knowledge. More specifically, we demonstrate that, even under substantial uncertainty about the decision maker’s HMM, the methods developed herein can devastate inference quality. Moreover, we also explore the effectiveness of the CME attack and its limitations with respect to instance size. We examine each  
problem-and-objective-function combination with this attack by utilizing a modestly sized HMM for tractable illustration. Whereas in-depth testing of the R\&S and APS solution approaches are performed in subsequent sections, this section also demonstrates the marginal effect of allocating additional computational resources for each attack. In so doing, we highlight that our attacks can expeditiously identify quality solutions.

Assume that Player $D$'s true HMM is defined by

\begin{gather*}
A_D =  \begin{bmatrix}
0.85 & 0.05 & 0.1 \\
0.05 & 0.9 & 0.05 \\
0.5 & 0.25 & 0.25 
\end{bmatrix}  , \\
B_D =  \begin{bmatrix}
0.699 & 0.05 & 0.1 & 0.05 & 0.1 & 0.001  \\
0.001 & 0.1 & 0.1 & 0.299 & 0.3 & 0.2 \\
0.1 & 0.2 & 0.1 & 0.2 & 0.1 & 0.3 
\end{bmatrix} , \\ 
\pi_D =  \begin{bmatrix}
0.5 & 0.3 & 0.2
\end{bmatrix},
\end{gather*}
\noindent such that there are three possible latent states and six possible emissions at each state. This precise parameterization is unknown to Player $Z$; however, we assume that the attacker's beliefs are correct in expectation (i.e., $\mathbbm{E}[A]=A_D$). The attacker is also assumed to believe that the probability of a successful attack on an observation is constant. In this manner, we define two distinct uncertainty levels upon which to test our attacks. The lower uncertainty condition is characterized by a Dirichlet precision (i.e., $\kappa$)\footnote{Recall that the precision of a Dirichlet distribution equals the sum of its parameters. These values are varied under the constraint that, in expectation, Player $Z$'s beliefs coincide with Player $D$'s true parameterization.} of 10,000 and a constant attack-success probability of 0.95. Conversely, the high uncertainty condition corresponds to a Dirichlet precision of 100 and an attack-success probability of 0.75.
Assuming $\mathcal{T}=\{1,...,5\}$, Player $Z$ observes the true emissions  $X=\{5, 4, 6, 4, 5\}$ and wishes to thwart the decision maker's inference in accordance with their own self-interest. 

We examine Player $Z$'s behavior in the state-attraction, state-repulsion\footnote{The state-attraction and -repulsion problems are, structurally speaking, the same problem. However, we solve both herein to explore how varying objective-function parameterizations may affect the attacker's solution. For brevity, we exclude such exploration of the remaining problems.}, distribution-disruption, and path-attraction problems under the two aforementioned uncertainty levels and varying ratios of $\nicefrac{w_1}{w_2}$. For the state-attraction problem, Player $Z$ desires Player $D$ to believe $Q_3=1$, whereas in the state-repulsion problem, Player $Z$ desires Player $D$ not to believe $Q_3=2$. 
In the distribution-disruption problem, Player $Z$ wants to maximally perturb Player $D$'s smoothing distribution at $t'=3$ while, in the path-attraction problem, Player $Z$ wishes the decision maker to infer $Q =\{3, 1, 1, 1, 3\}$ as the most probable sequence of hidden states. The attacker aims to maximize their expected utility in these settings; however, to highlight the effect of HMM corruption for each problem, distinct performance measures are reported. More specifically, performance measures for the state-attraction, state-repulsion, distribution-disruption, and path-attraction problems are respectively set as Player $D$'s subjective belief of $Q_3=1$, Player $D$'s subjective belief of $Q_3=2$, the Kullback Leibler divergence between the true and corrupted smoothing distributions, and the normalized Hamming distance\footnote{The normalized Hamming distance between two strings with the same length is the number of positions at which the corresponding symbols are different, divided by the strings' length.}.

We approximate the effect of $\hat{z}^*$ on these metrics for each problem-and-uncertainty combination as a function of $\nicefrac{w_1}{w_2}$. An $N$-value of 10,000 was utilized to estimate $f_1(\cdot)$ for each attack alternative and problem type. These estimates were subsequently used to estimate the expected utility of every attack for each $\nicefrac{w_1}{w_2}$. The attack alternatives that maximize the expected utility were selected as $\hat{z}^*$-variables. Once identified, the attack $\hat{z}^*$ was simulated $M$ times to generate an approximate distribution over the performance metrics. The value of $M$ was selected to ensure an appropriate level of precision for the mean performance in each problem. This required $M=5,000$ simulations for the state-attraction and distribution-disruption problems, but $M=1,000$  
sufficed for the state-repulsion and path-attraction problems. Within each 
problem, the associated $M$-value is kept constant across uncertainty levels to ensure proper comparisons. 

\begin{figure}[htbp!]
\begin{subfigure}[b]{0.4\textwidth}
    \centering
    \includegraphics[width=0.65\textwidth]{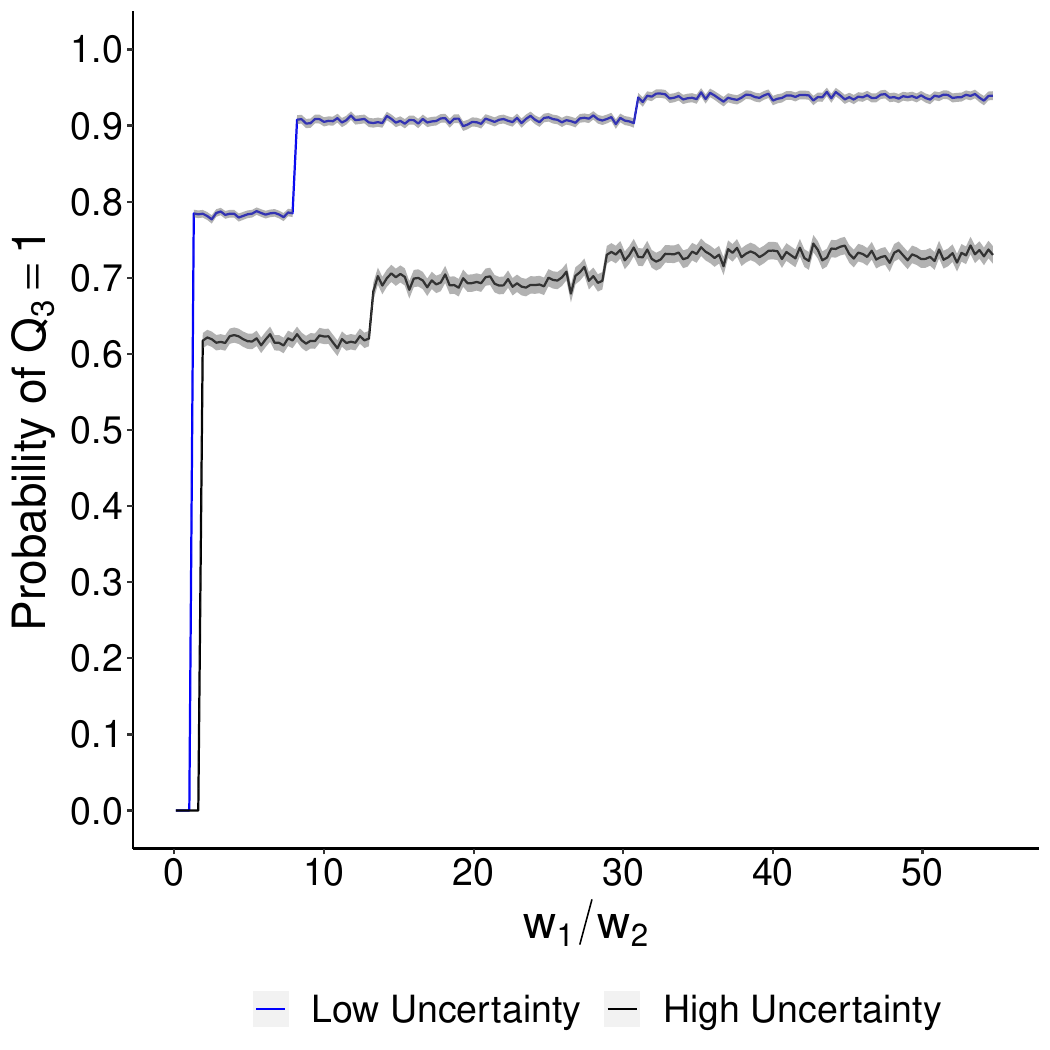}
    \caption{\centering State Attraction}
    \label{fig:att_ratio}
\end{subfigure} 
\begin{subfigure}[b]{0.4\textwidth}
    \centering
    \includegraphics[width=0.65\textwidth]{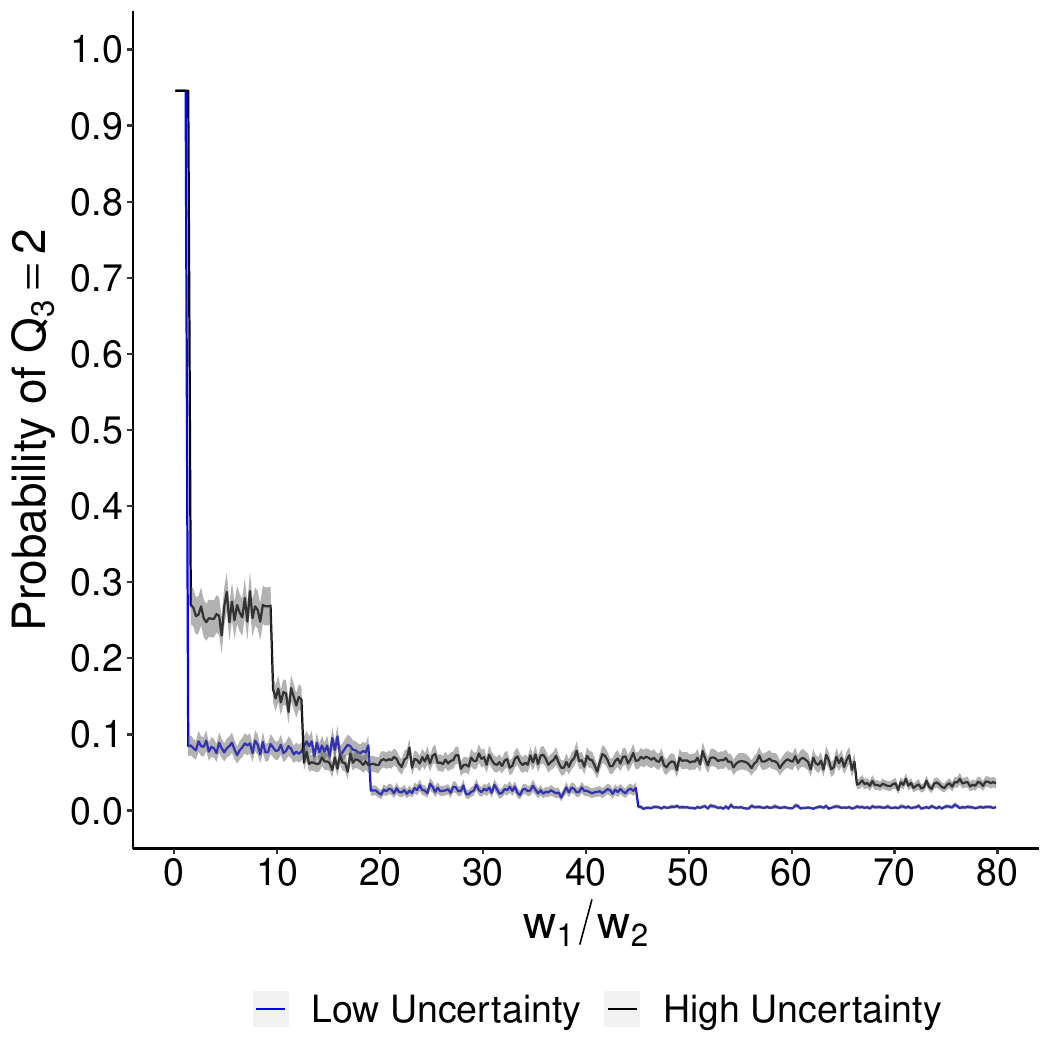}
    \caption{\centering State Repulsion}
    \label{fig:rep_ratio}
\end{subfigure}\\
\begin{subfigure}[b]{0.4\textwidth}
    \centering
    \includegraphics[width=0.65\textwidth]{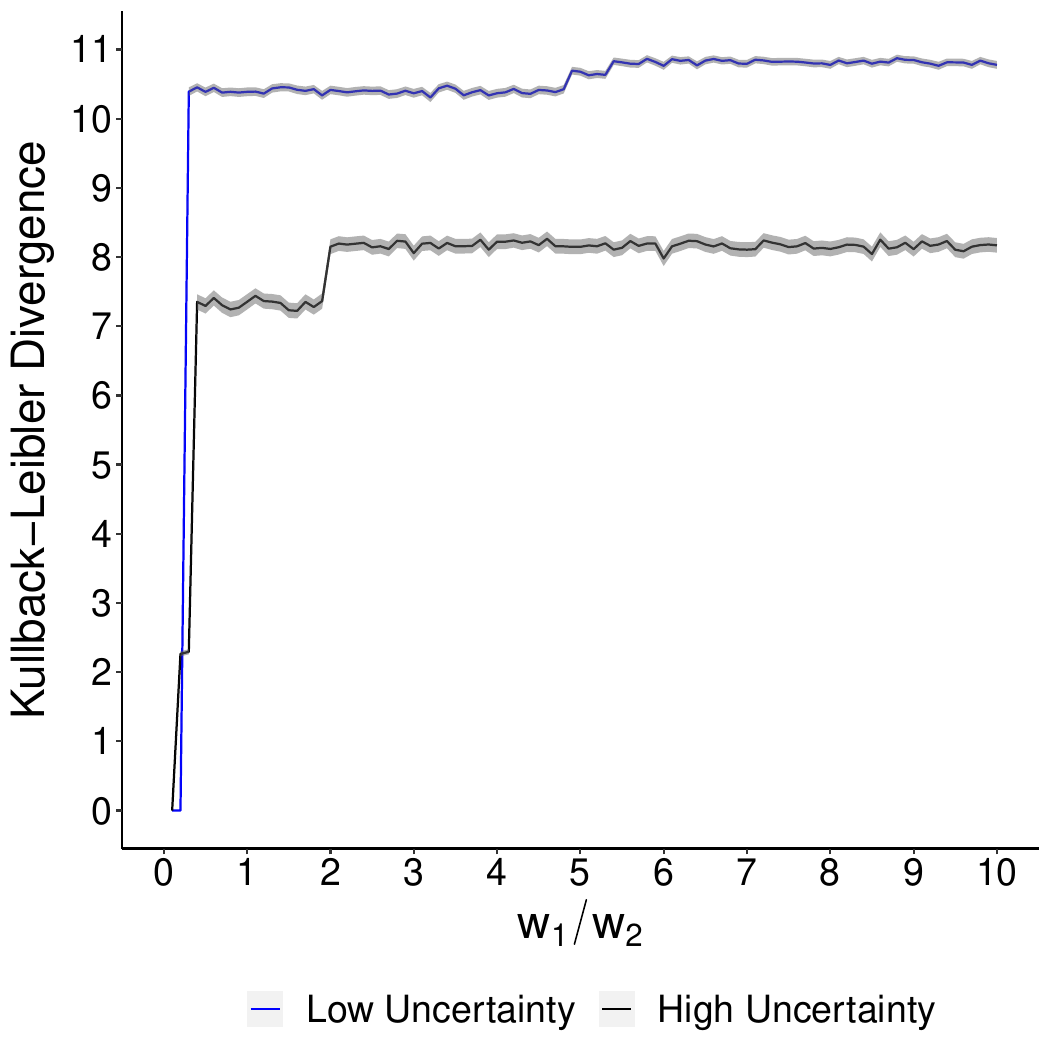}
    \caption{Distribution Disruption}
    \label{fig:dist_ratio}
\end{subfigure}
\begin{subfigure}[b]{0.4\textwidth}
    \centering
    \includegraphics[width=0.65\textwidth]{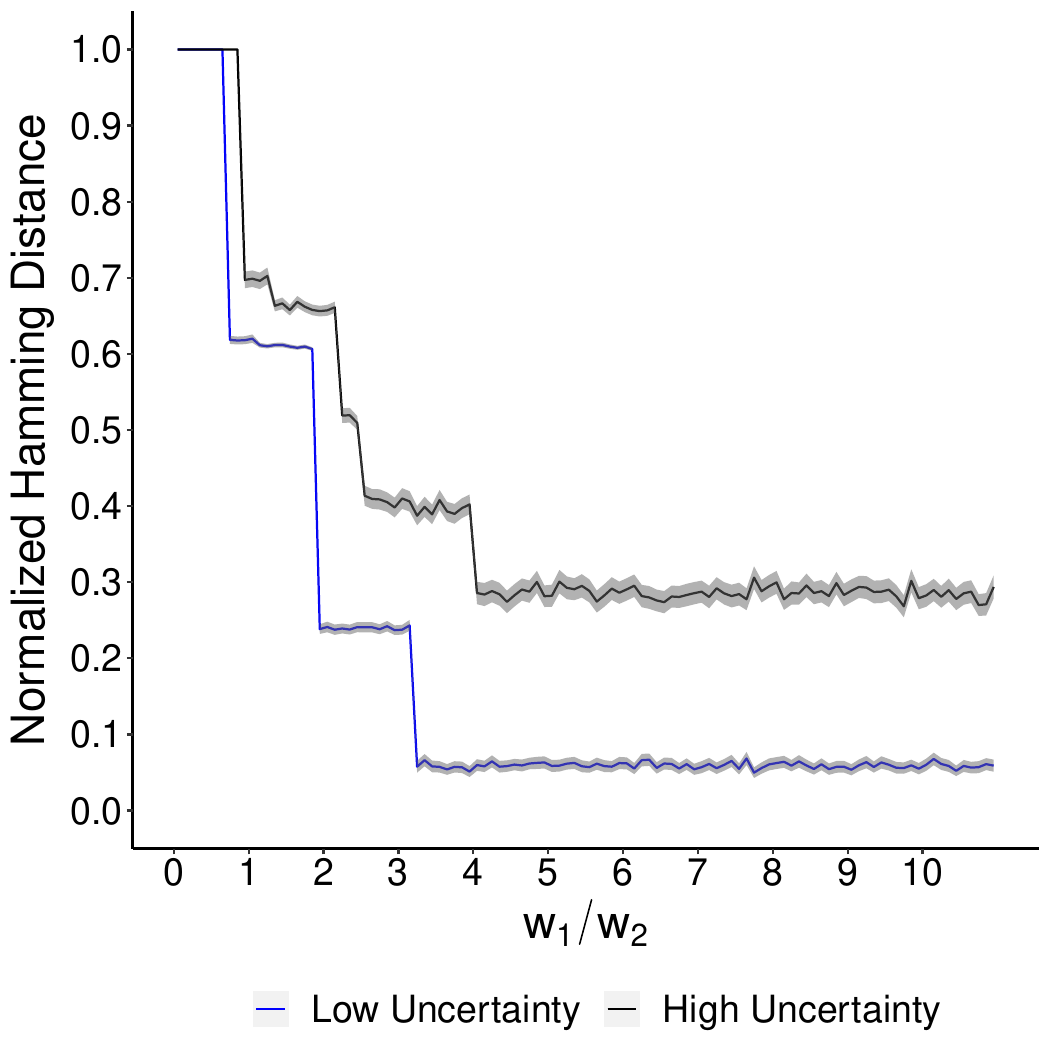}
    \caption{Path Attraction}
    \label{fig:path_ratio}
\end{subfigure} \\
\centering
\caption{Ratio Plots for each Problem-and-Uncertainty Pair}
\label{fig:ratios}
\end{figure}

Figure \ref{fig:ratios} summarizes the results of this testing for each problem-and-uncertainty pair. Within each plot, the mean performance measures over the $M$ samples are represented by a blue and black line for the low- and high-uncertainty levels, respectively; $\pm 2 \left(\nicefrac{s}{\sqrt{M}}\right)$ confidence regions are shaded grey. It can be observed that, in any problem when $\nicefrac{w_1}{w_2} \to 0$, the attacker determines the benefits of corrupting the data are outweighed by its costs. The attacker chooses to keep $\{x_t\}_{t\in \mathcal{T}}$ unchanged and, in so doing, concedes undesirable inference behavior by Player $D$. For example, in the state-repulsion problem, when $\nicefrac{w_1}{w_2} \to 0$, the probability of $Q_3 = 2$ is approximately 0.95, i.e., the probability inferred by the decision maker using the true observations.
However, as $\nicefrac{w_1}{w_2}$ increases and relative corruption costs decrease, this balance shifts and Player $Z$ begins to corrupt the data. Notably, across all plots, it is interesting to note that step-function-like behavior is induced. This behavior can also be observed by inspecting $\hat{z}^*$ at each $\nicefrac{w_1}{w_2}$ for every problem. That is, $\hat{z}^*$ tends to stay constant across an interval of ratios until the reward of attacking some $x_t$ is worth the penalty, at which point a new optimal attack is determined. The result of this behavior is that the expected value of the performance measures stays constant over the domain wherein $\hat{z}^*$ is constant.

Inspection of Figure \ref{fig:ratios} reveals that distinct values of $\nicefrac{w_1}{w_2}$ are necessary to induce modified attacker behavior. For example, in the state-attraction problems, Player $Z$ is incentivized to maximally perturb the observation vector at 
$\nicefrac{w_1}{w_2}\approx 30$, but this behavior is not induced in the path-attraction problems until $\nicefrac{w_1}{w_2} > 3$.
\
By inspecting each subfigure in Figure \ref{fig:ratios}, once can discern how the uncertainty levels affect the attacker's behavior. Notably, within each problem, the high-uncertainty level alters the range of $\nicefrac{w_1}{w_2}$-values in which a $\hat{z}^*$ is estimated to be optimal and, in expectation, tends to induce a less-preferable, maximally perturbed outcome. Likewise, tighter confidence regions about the attack's expected value can be derived under low uncertainty than high uncertainty.

The plots in Figure \ref{fig:ratios} are also interesting in that, once $\nicefrac{w_1}{w_2}$ is sufficiently far from zero, a dramatic increase in the performance measure of each problem is observed. The degree of improvement is problem dependent (e.g., compare the behavior of Subfigures \ref{fig:dist_ratio} and \ref{fig:path_ratio} near $\nicefrac{w_1}{w_2}=0$), but the pattern is consistent. Nevertheless, this behavior naturally leads one to inquire about the nature of the requisite attacks, i.e., the degree to which they perturb $\{x_t\}_{t\in \mathcal{T}}$. Fortunately, the size of the HMM examined in this section allows us to readily examine this in greater detail. 

\begin{figure}
\begin{subfigure}[b]{0.45\textwidth}
    \centering
    \includegraphics[width=0.65\textwidth]{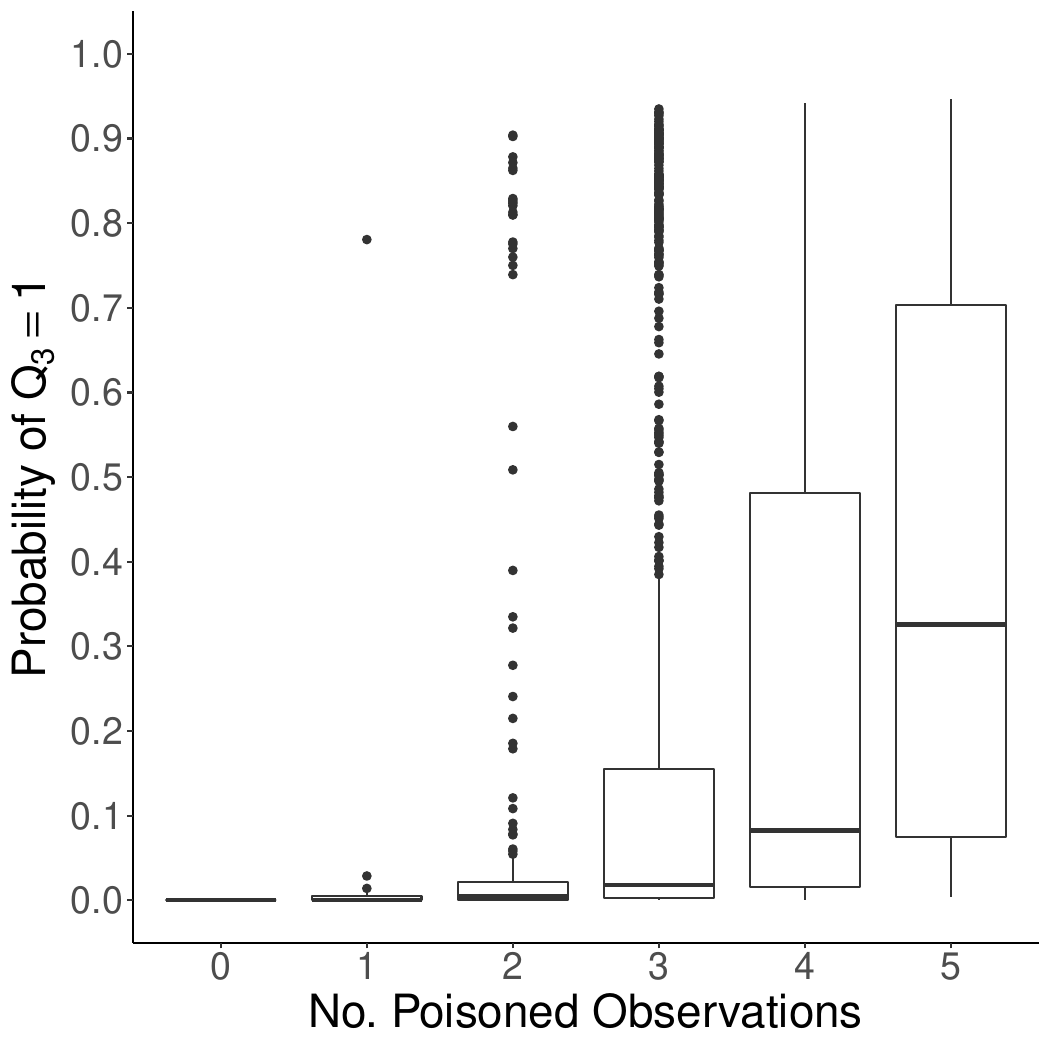}
    \caption{\centering State-Attraction, Low Uncertainty}
    \label{fig:att_low_box}
\end{subfigure} 
\begin{subfigure}[b]{0.45\textwidth}
    \centering
    \includegraphics[width=0.65\textwidth]{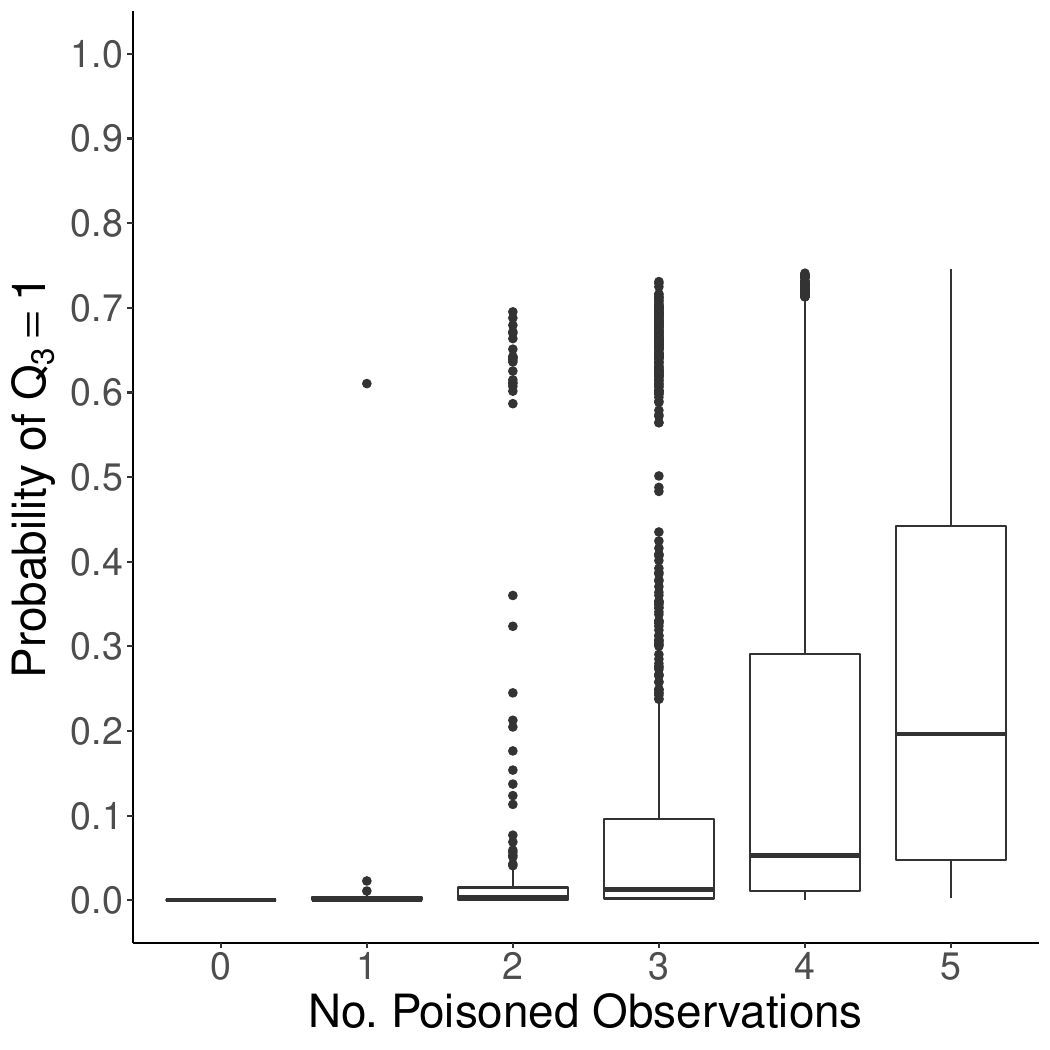}
    \caption{\centering State-Attraction, High Uncertainty}
    \label{fig:att_high_box}
\end{subfigure}\\
\begin{subfigure}[b]{0.45\textwidth}
    \centering
    \includegraphics[width=0.65\textwidth]{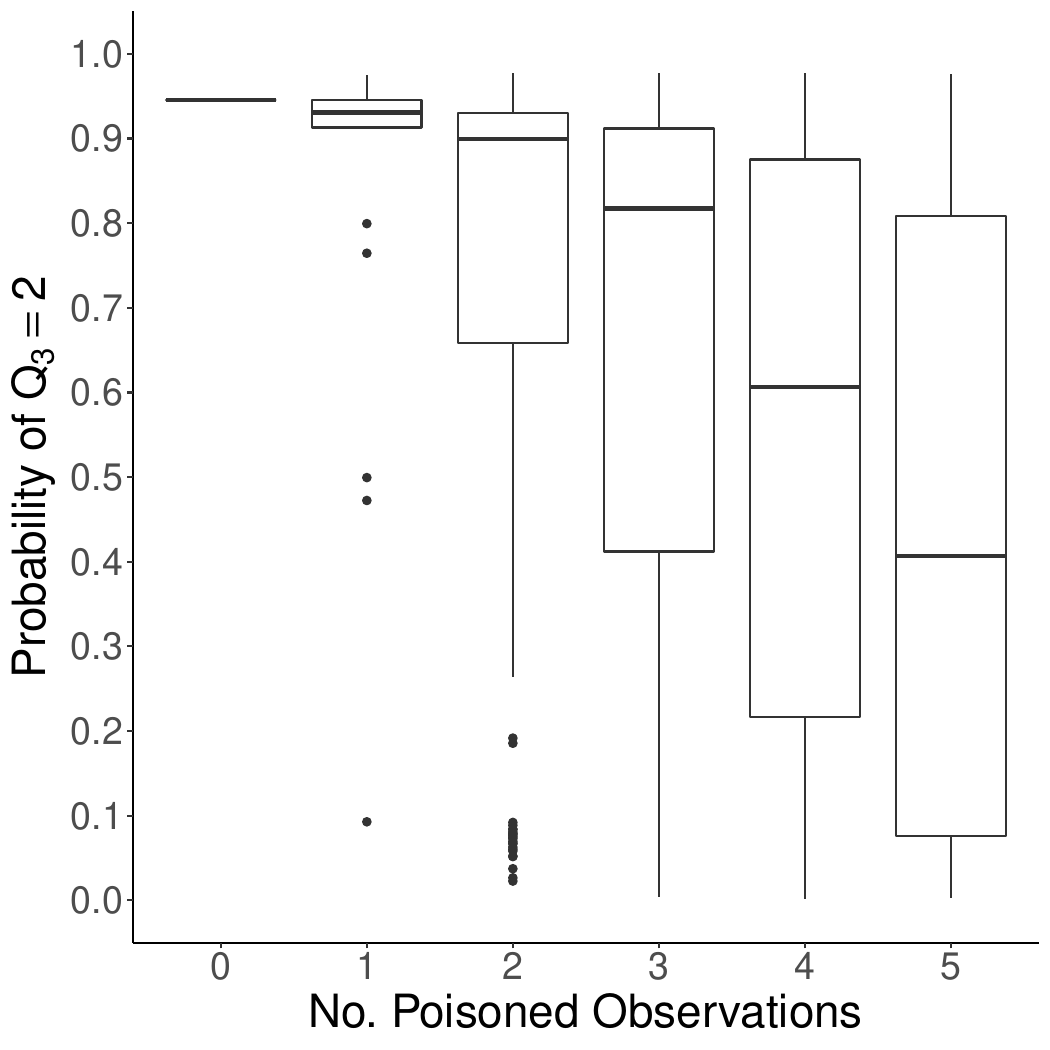}
    \caption{State-Repulsion, Low Uncertainty}
    \label{fig:rep_low_box}
\end{subfigure}
\begin{subfigure}[b]{0.45\textwidth}
    \centering
    \includegraphics[width=0.65\textwidth]{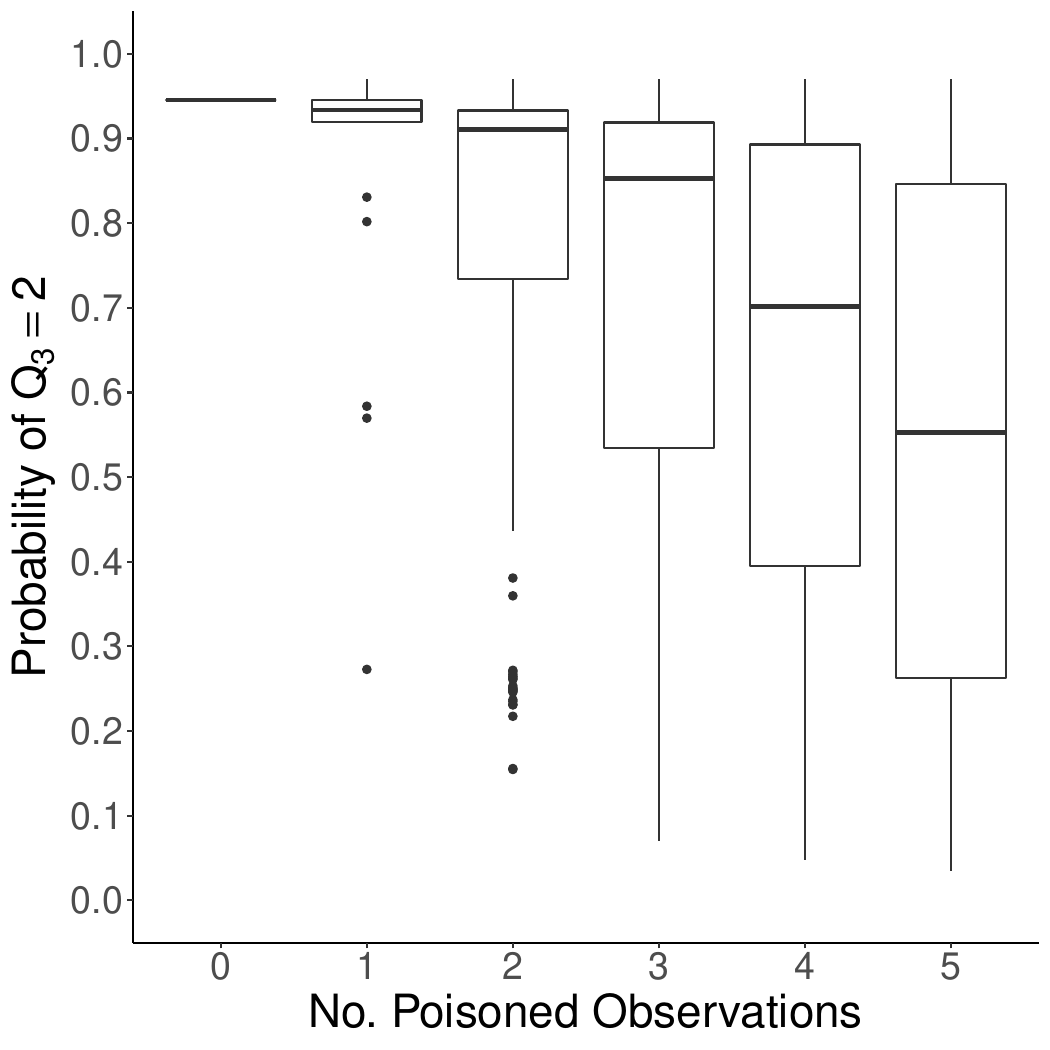}
    \caption{State-Repulsion, High Uncertainty}
    \label{fig:rep_high_box}
\end{subfigure} \\
\begin{subfigure}[b]{0.45\textwidth}
    \centering
    \includegraphics[width=0.65\textwidth]{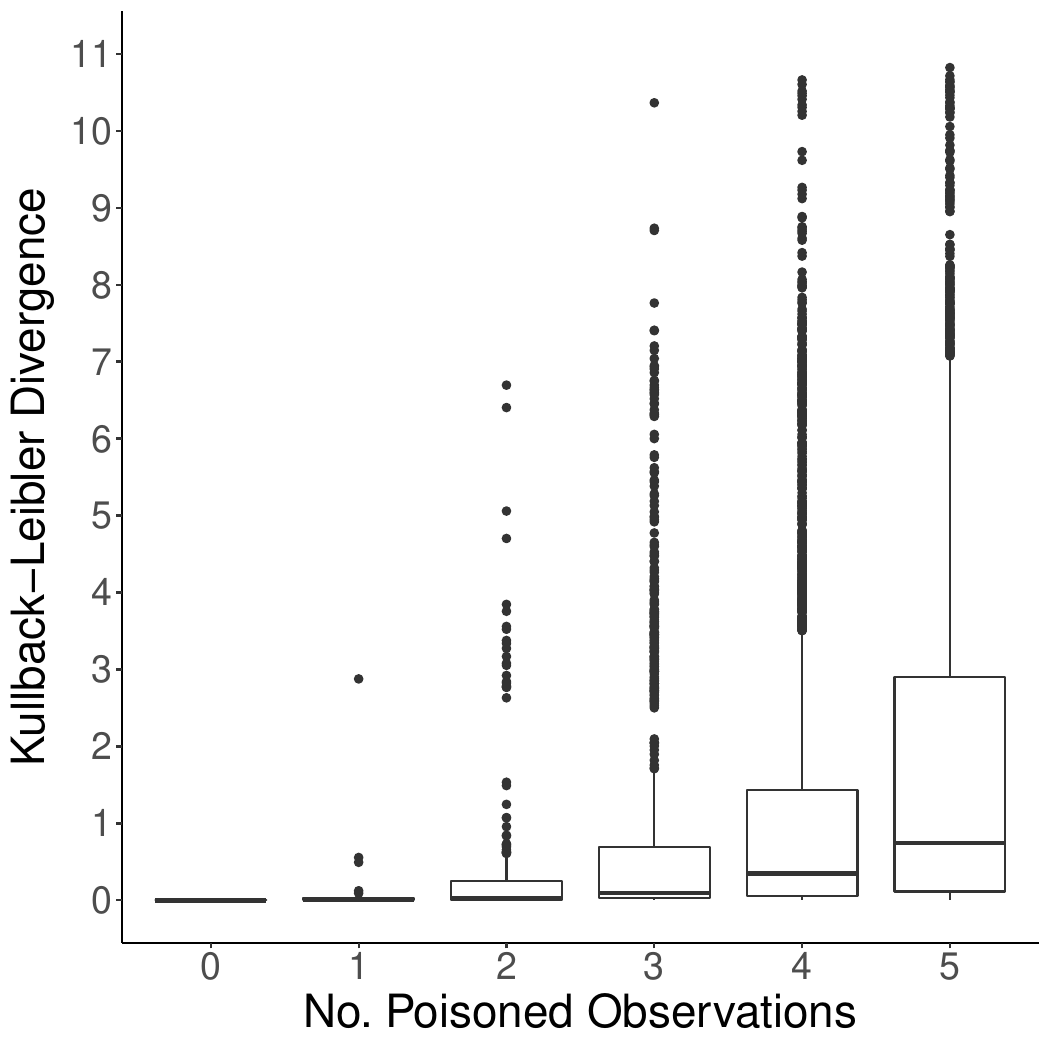}
    \caption{Distribution-Disruption, Low Uncertainty}
    \label{fig:dd_low_box}
\end{subfigure}
\begin{subfigure}[b]{0.45\textwidth}
    \centering
    \includegraphics[width=0.65\textwidth]{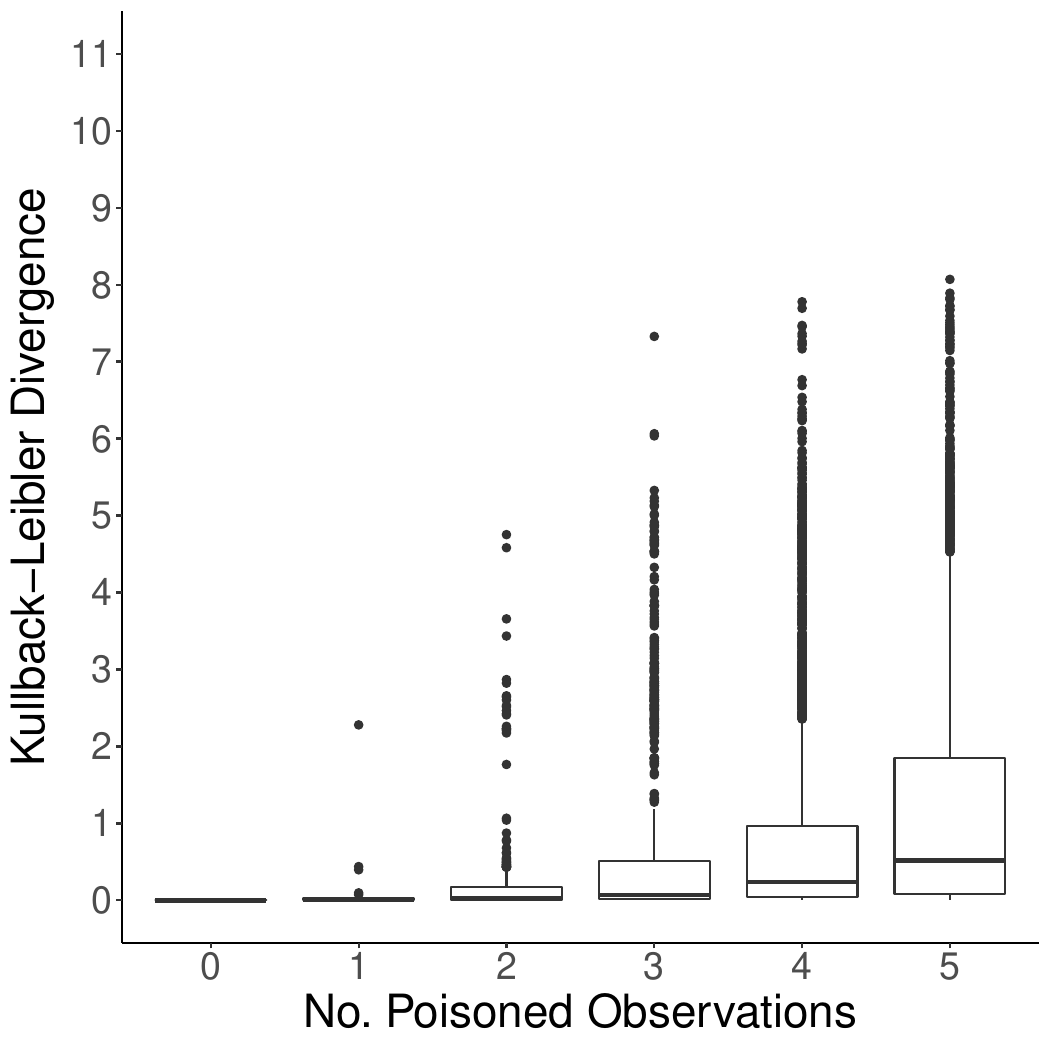}
    \caption{Distribution-Disruption, High Uncertainty}
    \label{fig:dd_high_box}
\end{subfigure} \\
\begin{subfigure}[b]{0.45\textwidth}
    \centering
    \includegraphics[width=0.65\textwidth]{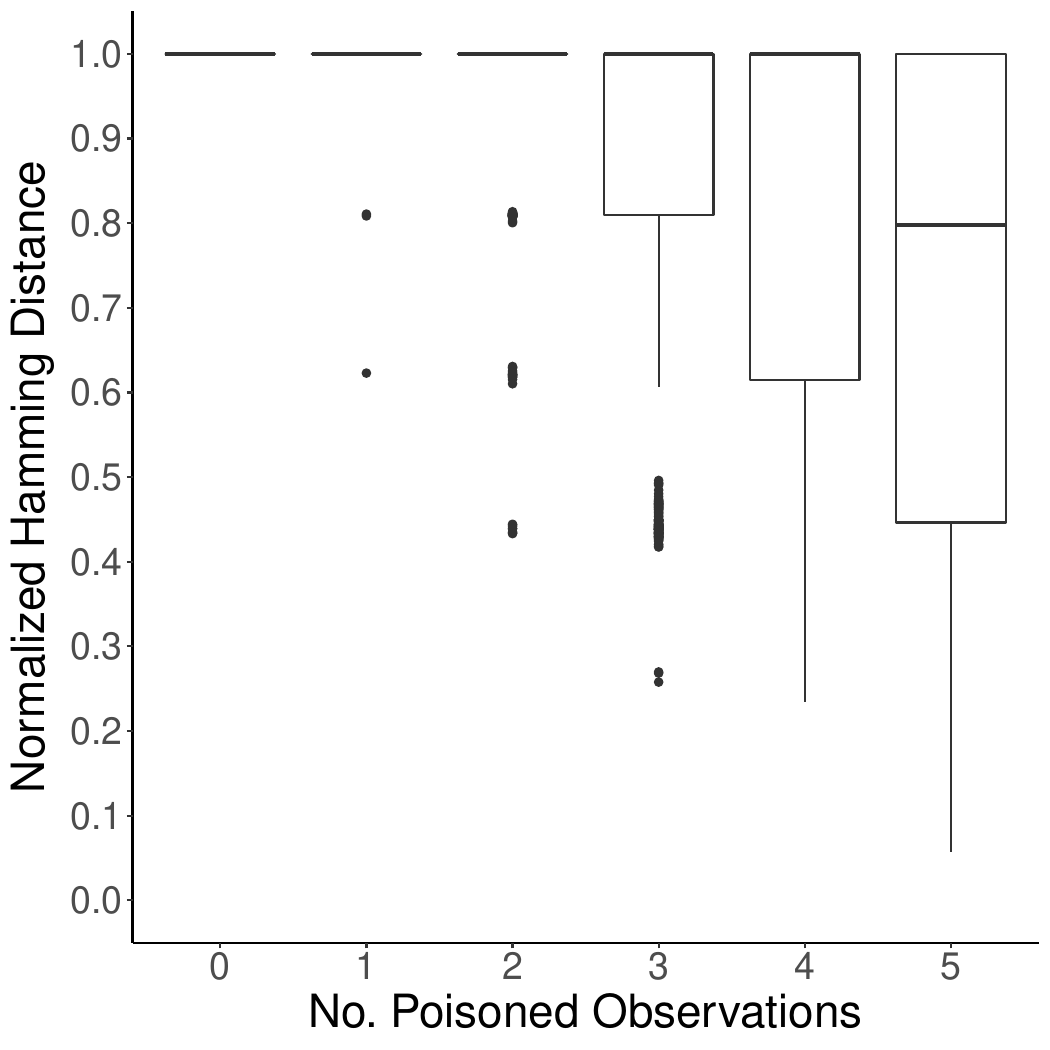}
    \caption{Path-Attraction, Low Uncertainty}
    \label{fig:pd_low_box}
\end{subfigure}
\begin{subfigure}[b]{0.45\textwidth}
    \centering
    \includegraphics[width=0.65\textwidth]{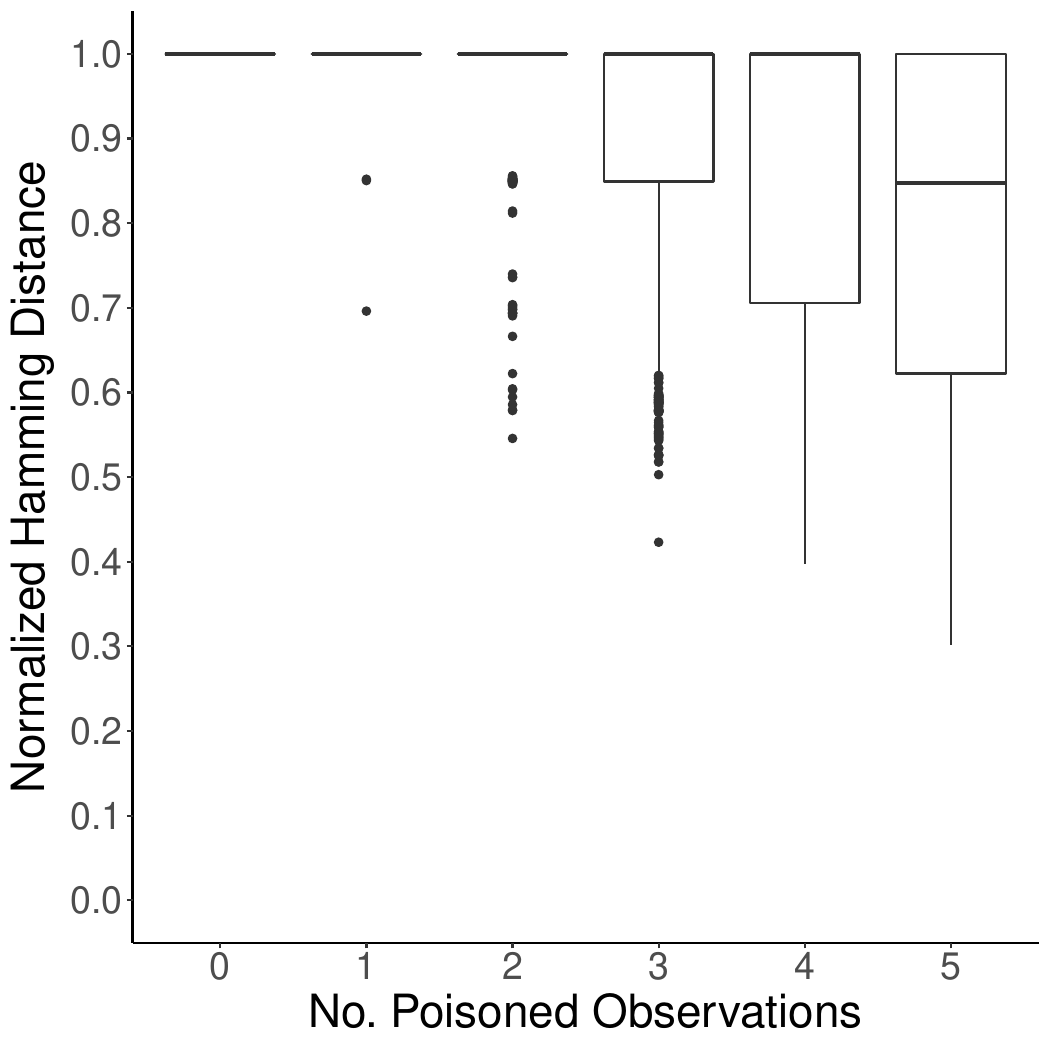}
    \caption{Path-Attraction, High Uncertainty}
    \label{fig:pd_high_box}
\end{subfigure}
\centering
\caption{Mean Attack Efficacy by Number of Perturbed Observations}
\label{fig:boxplots_5_1}
\end{figure}

We perform additional analysis on the attacker's problems by enumerating the effects of every $z \in \mathcal{Z}$; $M$ simulations of each attack $z$ are performed to estimate its mean performance for the aforementioned problem types. The box plots in Figure \ref{fig:boxplots_5_1} examine to what degree attacks of differing perturbation levels  (i.e., the number of corrupted observations) affect the decision maker's inference. Examination of these plots yields noteworthy insights. 
The attacker could considerably disrupt Player $D$'s inference by corrupting only a single observation in some cases, assuming they have the means to readily identify it. For example, in the case of low uncertainty, corrupting one observation could change Player $D$'s subjective probability from $\approx 0$ to $0.8$ in the state-attraction problem while a similar change from $\approx 0.95$ to $0.1$ is observed in the state-repulsion problem. Such modest attacks are not quite as successful in the distribution-disruption or path-attraction problems, but high-quality results can still be achieved without corrupting the entirety of $\{x_t\}_{t\in \mathcal{T}}$. Moreover, it can also be observed in the variability of each subfigure that high-perturbation attacks are not necessarily effective; in fact, the variability in an attack's success increases with the number of corrupted observations. 
Although these results collectively highlight the disruption that HMM corruption can induce on a decision maker's inference, they also emphasize the difficulty of the attacker's problems. This is most apparent by considering the medians and inter-quartile ranges in Figure \ref{fig:boxplots_5_1}. Effective attacks that limit the number of corrupted observation are the exception, not the rule. Conversely, blindly corrupting a large number of observations has no guarantee of success; such attacks have better median performance but are highly variable. 

Finally, perhaps the most noteworthy result in this section pertains to the performance of the CME algorithm. Despite the difficulty of the examined problems, this solution methodology was readily able to identify high-quality solutions. Furthermore, it is also worth emphasizing that, whereas the attacks are built under both aleatoric and epistemic uncertainty (i.e., the unknown attack success and HMM parameters, respectively), the evaluations provided herein are performed under aleatoric uncertainty only. This fact further emphasizes the utility of our methodology. Despite not knowing the outcome of an attack or the decision maker's true HMM at design time, our attacks were able to substantially thwart Player $D$'s inferences. 

Such a result bodes well for the real-world applicability of such corruption attacks. The attacker is unlikely to know the true HMM, but the methods presented herein allow  
to identify high-quality attacks despite this limitation. Unfortunately, although the CME method is effective, additional experimentation found it too computationally burdensome for larger instances. Given some $\nicefrac{w_1}{w_2}$, the problems examined within this section can generally be solved within 30-55 minutes using the aforementioned hardware but, for larger-sized HMMs having a $\mathcal{T}$ of greater cardinality, the required computational effort rapidly increases.
This motivates the utilization of other methods, e.g., the R\&S and APS methods discussed in the next two subsections.

\subsection{Effect of HMM Structure on Solution Method Performance} \label{secStrucTest}

This section focuses on the effect of the HMM structure on the R\&S and APS methods. Following the best practices set forth by \citet{coffin2000statistical}, we consider the $2^{3-1}_{III}$ design presented in Table \ref{tab:hmm_size3_new} upon which to test algorithm performance. The distributions defining the decision maker's true HMM parameters (e.g., $A_D$) are generated randomly via a Dirichlet distribution with all concentration parameters set to one. The true observations are generated randomly from a discrete uniform distribution with support $\{1,2,... |X|\}$, whereas $\lambda= 0.95$ and $\kappa = 10,000$ for each. Table \ref{tab:target_hmm_Struct_new} depicts the attacker's goals for each problem, and the objective function weight ratios, $\nicefrac{w_1}{w_2}$, are selected to correspond with a conservative attacker who is unwilling to corrupt all observations. The complete parameterizations are available online. \footnote{Available at https://github.com/roinaveiro/corrupting\_hmms}

\begin{table}[H]
\centering
\caption{ $2^{3-1}_{III}$ Design For HMM Structure Testing}
\label{tab:hmm_size3_new}
\resizebox{0.35\textwidth}{!}{
\begin{tabular}{cccc}
\hline
Design Point & $|Q|$ & $|X|$ & $|T|$ \\ \hline
1           & 30    & 30  & 30    \\ 
2           & 10    & 10  & 30    \\ 
3           & 10    & 30  & 10    \\ 
4           & 30    & 10  & 10    \\ \hline
\end{tabular}
}
\end{table}

\begin{table}[H]
\centering
\caption{Attacker's Objective Function Parameters by Problem and Design Point}
\label{tab:target_hmm_Struct_new}
\resizebox{\textwidth}{!}{
\begin{tabular}{cccccc}
\hline
\multirow{2}{*}{Problem}& \multirow{2}{*}{$\nicefrac{w_1}{w_2}$} & \multicolumn{4}{c}{Design Point} \\  \cmidrule(lr){3-6}
   & & 1  & 2  & 3  & 4  \\ \hline
State-Attraction & 20        & \{$t'=$ 25, $i' =$ 4\} & \{$t'=$21, $i' =$9\} & \{$t'=$9, $i'=$7\}  & \{$t'=$9, $i'=$22\} \\ 
State-Repulsion & 15        & \{$t' =$ 23, $i' =$ 9\} & \{$t'=$29, $i' = $ 4\} & \{$t' = $5, $i' = $6\} & \{$t'=$9, $i'=$12\} \\ 
Distribution-Disruption & 3 & $t' = $ 23 & $t'=$29 & $t'=$5 & $t'=$9 \\ 
Path-Attraction    & 2.55      &  \multicolumn{4}{c}{$\{0\}_{\forall t \in\mathcal{T}}$} \\ \hline
\end{tabular}
}
\end{table}

Two variants of each of the R\&S and APS methods are tested against the RME algorithms; the CME approach is ignored because empirical testing found it to be overly encumbered for instances of this size. The two R\&S methods are henceforth referred to as R\&S-A and R\&S-B. Both utilize a multi-layer perceptron neural network as $\hat{\mu}(\cdot)$ to regress the attack's expected utility, but differ in their use of Monte-Carlo tree search and simulated annealing to identify a greedy action, respectively. Further details on the implementation of these methods, as well the hyperparameter tuning that informed them, is provided in \ref{appendix_RandS}. Likewise, the APS techniques are referenced as APS-A and APS-B; the approaches have annealing schedule of $\lbrace \mathcal{H}_n \rbrace_{n=1}^\infty$ and $\lbrace \mathcal{H}_n \rbrace_{n=500}^\infty$, respectively. These schedules result in APS-A exploring the solution space more thoroughly than APS-B which instead favors exploitation.

We explore the efficacy of these attacks both in terms of solution quality and computational effort. This is accomplished by allowing each algorithm to operate for $\{15,30,45,...,1200\}$ seconds on every problem instance; the expected utility of the output attack at each time limit is calculated. Due to the stochastic nature of the attacks and the solution approaches, this procedure is replicated 10 times for each algorithm; mean expected utilities across these repetitions plus/minus two standard deviations are reported.

\begin{figure}[H]
    \centering
    \includegraphics[width=0.8\textwidth]{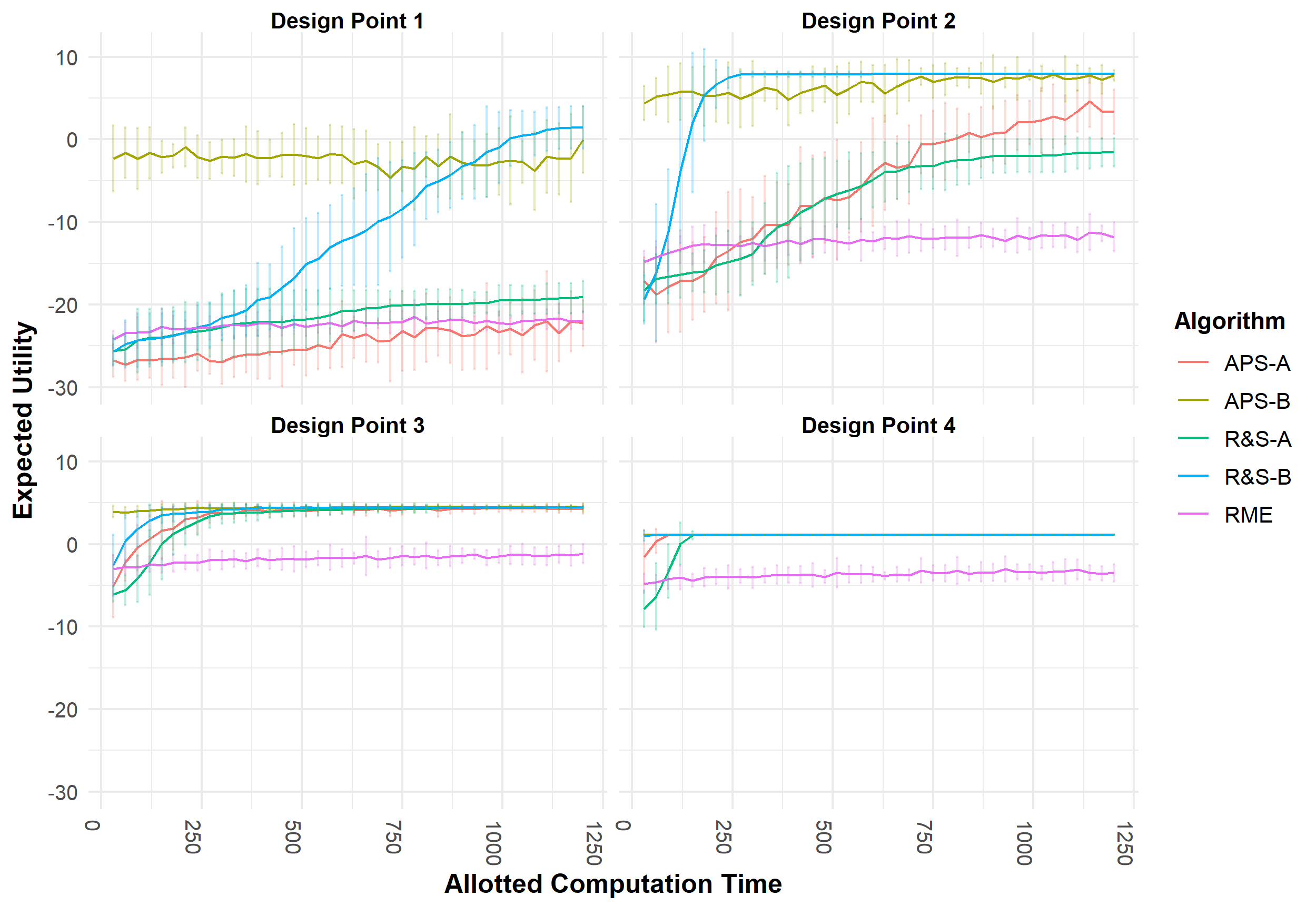}
    \caption{Evolution of Path-Attraction Expected Utilities over time by Algorithm}
    \label{fig:main_pd3_new}
\end{figure}

Several trends emerged from this analysis across each problem type. This behavior is typified within the path-attraction instances as provided in Figure \ref{fig:main_pd3_new}. Notably, APS-B and R\&S-B converge toward attacks having comparable expected utilities across all instances and, although not a firm rule, APS-B tends to do so quicker than R\&S-B. Alternatively, APS-A and R\&S-A are more variable. For instances defined by smaller $|\mathcal{T}|$ (i.e., Design Points 3 and 4) they also converge toward high-quality solutions but, for larger $|\mathcal{T}|$-values, they do not identify valuable attacks.
The computational time required to identify such solutions is variable as well. In juxtaposition, RME consistently identifies lower-quality attacks in every instance. Although not depicted herein, similar trends are also apparent across the other three problem types. 

Despite their global similarities, it is interesting to note that the algorithms vary systematically in the type of attacks they identify. Table \ref{tab:struct_attacks_allprobs} compares and contrasts  each algorithm  with respect to the performance of their final attacks, i.e., those identified at 1200 seconds, across all problem-and-design-point pairs. Performance is based on the respective objective function of the problem of interest. To account for the stochasticity in attack success, solution quality was estimated by simulating their effects 500 times for the state-attraction and distribution-disruption instances, and 100 times for the state-repulsion and path-attraction instances; averages are reported across all repetitions.
The impact measure is problem-specific and refers to damage caused to the decision maker's inference, whereas $\Delta$ refers to the total number of observations attacked. The greatest impact and the least $\nicefrac{\Delta}{|\mathcal{T}|}$
achieved by the algorithms are bolded for each problem-and-design-point pair. 

Inspection of this table reveals additional patterns obfuscated by the expected-utility calculations. APS-B and R\&S-B are better able to balance the attacker's multi-objective utility function; however, in so doing, they often find less damaging attacks to the decision maker's inference than their competitors. In particular, of the 16 problem-and-design-point pairs, the APS-B and R\&S-B algorithms identified the most impactful attack two and four times, respectively. The expected impact varied substantially about problem types as well. R\&S-B never identified the most impactful distribution-disruption attack; APS-B never did so for the state-attraction problem. However, the APS-B attacks appear to perturb relatively few perturbations in comparison to their impact. For specific problem-and-design-point pairs, the other three methods identified more impactful attacks. This success, however, was generally counterbalanced by higher rates of data perturbation. The APS-A approach is an exception to this rule in that it at times acts too conservatively by perturbing relatively few observations. 

\begin{table}[htb]
\centering
\caption{Mean Impact and Proportion of Observations Attacked (Structure Testing)}
\label{tab:struct_attacks_allprobs}
\resizebox{0.8\textwidth}{!}{
\begin{tabular}{cccccccccc}
\hline
\multirow{2}{*}{Problem} & \multirow{2}{*}{Algorithm} & \multicolumn{2}{c}{Design Pt. 1}                    & \multicolumn{2}{c}{Design Pt. 2}                    & \multicolumn{2}{c}{Design Pt. 3}                    & \multicolumn{2}{c}{Design Pt. 4}                   \\ \cmidrule(lr){3-4} \cmidrule(lr){5-6}  \cmidrule(lr){7-8}  \cmidrule(lr){9-10} 
  &                  & Impact & $\nicefrac{\Delta}{|\mathcal{T}|}$ & Impact & $\nicefrac{\Delta}{|\mathcal{T}|}$ & Impact & $\nicefrac{\Delta}{|\mathcal{T}|}$ & Impact & $\nicefrac{\Delta}{|\mathcal{T}|}$ \\ \hline
 & APS-A                  & 0.144  & 0.583                           & 0.040  & 0.030                           & 0.635  & 0.310                           & 0.124  & 0.130                           \\ 
\multirow{2}{*}{State}& APS-B                  & 0.136  & \textbf{0.043}                           & 0.009  & \textbf{0.007}                           & 0.636  & \textbf{0.300}                           & 0.120  & \textbf{0.120}                           \\ 
\multirow{2}{*}{Att.$^1$}& R\&S-A                    & 0.169  & 0.773                           & 0.038  & 0.517                           & \textbf{0.637}  & 0.420                           & \textbf{0.128}  & 0.280                           \\ 
& R\&S-B                      & \textbf{0.179}  & 0.327                           & 0.012  & 0.010                           & 0.627  & 0.340                           & 0.125  & 0.130                           \\ 
& RME                     & 0.125  & 0.830                           & \textbf{0.074}  & 0.687                           & 0.605  & 0.960                           & 0.127  & 0.580                           \\  \hline
& APS-A                   & 0.173  & 0.540                           & 0.502  & \textbf{0.033}                           & 0.568  & \textbf{0.100}                           & 0.168  & \textbf{0.100}                           \\ 
\multirow{2}{*}{State}& APS-B                   & 0.172  & \textbf{0.037}                           & 0.509  & \textbf{0.033}                           & 0.563  & \textbf{0.100}                           & \textbf{0.170}  & \textbf{0.100}                           \\ 
\multirow{2}{*}{Rep.$^2$}& R\&S-A                   & 0.174  & 0.740                           & 0.500  & 0.490                           & 0.566  & \textbf{0.100}                           & 0.166  & 0.180                           \\ 
& R\&S-B                      & \textbf{0.177}  & 0.163                           & \textbf{0.515}  & 0.047                           & 0.566  & 0.140                           & 0.169  & \textbf{0.100}                           \\ 
& RME                     & 0.167  & 0.810                           & 0.493  & 0.680                           & \textbf{0.577}  & 0.710                           & 0.167  & 0.520                           \\ \hline
& APS-A                   & 2.221  & 0.653                           & 2.202  & \textbf{0.033}                           & \textbf{1.445}  & 0.210                           & \textbf{1.520}  &\textbf{0.100}                           \\ 
\multirow{2}{*}{Dist.}& APS-B                   & 2.140  & \textbf{0.033}                           & 2.213  & \textbf{0.033}                           & 1.435  & \textbf{0.200}                           & 1.518  & \textbf{0.100}                           \\ 
\multirow{2}{*}{Disrupt.$^3$}&R\&S-A                    & \textbf{2.283}  & 0.747                           & \textbf{2.297}  & 0.510                           & 1.436  & 0.370                           & 1.459  & 0.200                           \\
& R\&S-B                      & 2.206  & 0.453                           & 2.279  & 0.327                           & 1.426  & 0.290                           & 1.511  & \textbf{0.100}                           \\ 
& RME                     & 2.203  & 0.877                           & 2.192  & 0.687                           & 1.270  & 0.830                           & 1.517  & 0.550                           \\ \hline
& APS-A                   & 0.980  & 0.817                           & \textbf{0.851}  & 0.190                           & \textbf{0.612}  & 0.340                           & \textbf{0.900}  & \textbf{0.000}                           \\ 
\multirow{2}{*}{Path} & APS-B                   & 0.940  & 0.080                           & 0.860  & \textbf{0.010}                           & 0.707  & \textbf{0.280}                           & \textbf{0.900}  & \textbf{0.000}                           \\ 
\multirow{2}{*}{Att.$^4$} & R\&S-A                    & 0.962  & 0.740                           & 0.857  & 0.483                           & 0.648  & 0.520                           & 0.921  & 0.150                           \\ 
& R\&S-B                    & \textbf{0.930}  & \textbf{0.027}                           & 0.867  & 0.017                           & 0.620  & 0.360                           & \textbf{0.900}  & \textbf{0.000}                           \\ 
& RME                     & 0.972  & 0.813                           & 0.897  & 0.690                           & 0.718  & 0.700                           & 0.923  & 0.460                           \\ \hline
\multicolumn{10}{l}{\scriptsize $^1$ Impact is the difference between perturbed and unperturbed data of the probability of $i'$ at $t'$  } \\ [-1ex] 
\multicolumn{10}{l}{\scriptsize $^2$ Impact is the difference between unperturbed and perturbed data of the probability of $i'$ at $t'$  } \\ [-1ex]
\multicolumn{10}{l}{\scriptsize $^3$ Impact is the KL divergence between state distributions under unperturbed and perturbed data} \\ [-1ex] 
\multicolumn{10}{l}{\scriptsize $^4$ Impact is the Normalized-Hamming distance of most-likely state sequences under perturbed data and attacker's goal} 
\end{tabular}
}

\end{table}

Collectively, these results illustrate that the APS and R\&S algorithms can effectively thwart larger-scale HMM instances. The APS-B and R\&S-B configurations were the most effective for the instances explored herein, but their performance varied. This variation was especially apparent across $|\mathcal{T}|$-values, a fact deriving from the exponential effect $|\mathcal{T}|$ has on the attack space's cardinality. Nevertheless, many of the results discussed herein are specific to the parameterization utilized within this section. Alternative uncertainty structures about the decision maker's HMM may also play a significant role in the algorithm's efficacy. Exploring such dynamics is the focus of the next subsection.

\subsection{Effect of Uncertainty on Solution Method Performance} \label{secUncTest}

The same algorithm variants studied in the previous section are utilized herein to explore the effect of uncertainty on their performances. The HMM size is fixed at $|\mathcal{Q}|$, $|\mathcal{X}|$, and $|\mathcal{T}| = 20$; however, the attacker's beliefs and the attack's success probability is varied via the $2^2$ factorial design provided in Table \ref{tab:unc_case_new}. The decision maker's true HMM parameters are built in a similar manner to Section \ref{secStrucTest} sampling from Dirichlet distributions; the true observations are likewise built by sampling from discrete uniform distributions. The $\nicefrac{w_1}{w_2}$-values for each problem are maintained constant from Section \ref{secStrucTest}, as is the attacker's goal in the path-attraction problem. However, their goals in the state-attraction, state-repulsion and distribution-disruption instances correspond to $\{t'=17, i'=4\}$, $\{t'=16, i'=7\}$, and $t'=16$, respectively. The complete parameterization of these instances is available online\footnote{Available at https://github.com/roinaveiro/corrupting\_hmms}. 

\begin{table}[H]
\centering
\caption{$2^2$ Full Factorial Design for Uncertainty Testing}
\label{tab:unc_case_new}
\resizebox{0.4\textwidth}{!}{
\begin{tabular}{ccccc}
\hline
\multirow{2}{*}{Parameter}& \multicolumn{4}{c}{Design Points}\\ \cmidrule(lr){2-5}
 & 1 & 2 & 3& 4 \\ \hline
$\lambda$    &     0.95     & 0.95 & 0.75 & 0.75      \\ 
$\kappa$    &  10,000    & 100 & 10,000 & 100       \\ \hline
\end{tabular}
}
\end{table}

\begin{figure}[htbp!]
    \centering
    \includegraphics[width=0.8\textwidth]{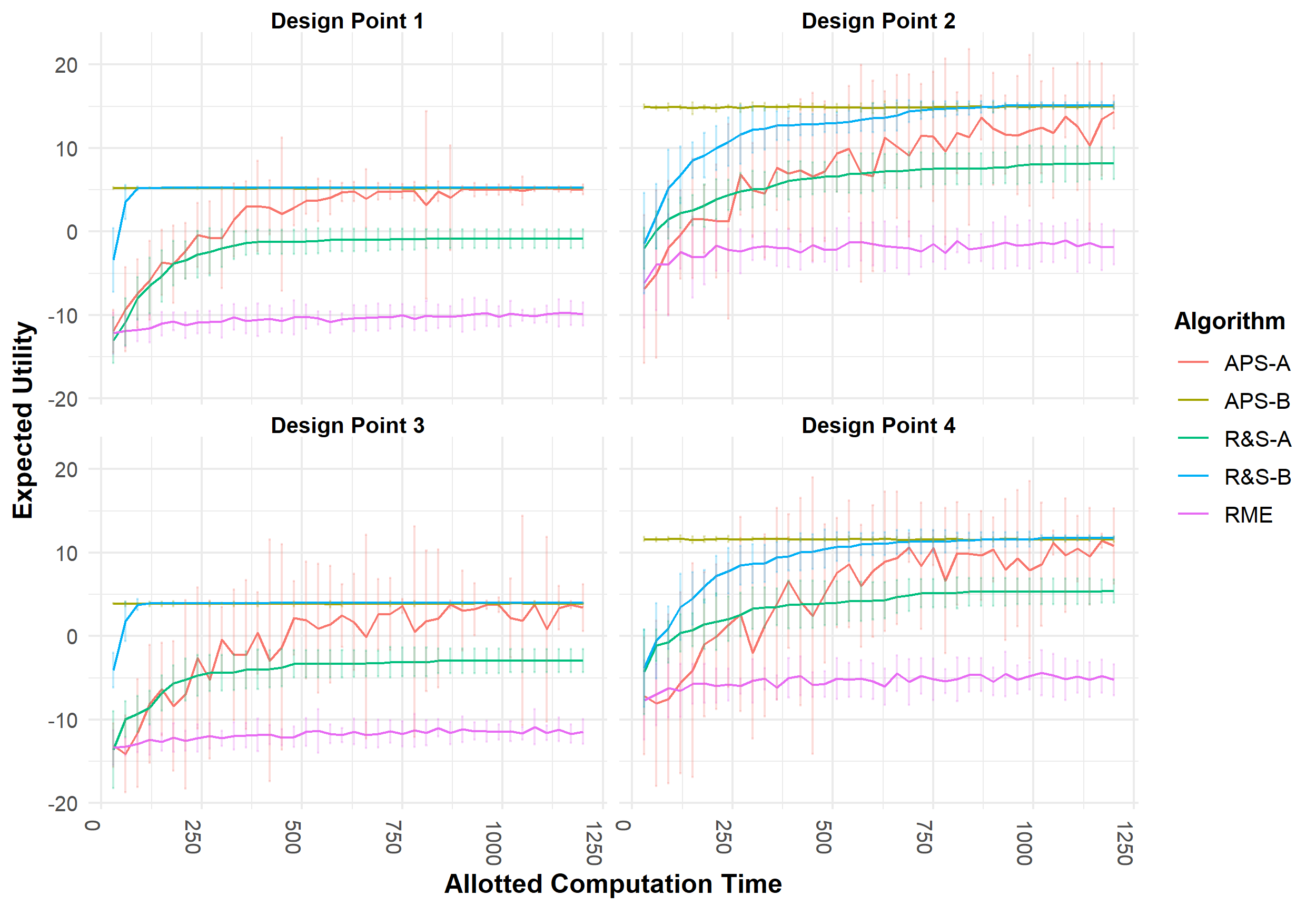}
    \caption{Temporal Evolution of Distribution-Disruption Expected Utility by Algorithm}
    \label{fig:main_distdisrupt_new}
\end{figure}

Testing that is similar to the previous sub-section was conducted on each algorithm for every problem-and-design-point pair. Figure \ref{fig:main_distdisrupt_new} provides the algorithms' performance in relation to the allocated computation time for the distribution-disruption problem. These results bear resemblance to those of the previous sub-section in the APS-B and R\&S-B tend to find comparable solutions, but with APS-B doing so more expeditiously. The RME method again underperforms with respect to its peers. Nevertheless, the performance of  APS-A is distinct in these instances from the previous section. It tends to converge toward comparable solutions to that found by APS-B and R\&S-B; however, the variability about its expected utility is great, indicating highly variable performance. As with the previous sub-section, similar trends are also apparent across the other three problem types.   

Table \ref{tab:unc_attacks_allprobs} depicts the mean performance of each algorithm for the final attacks across the four design points. These values are calculated by averaging the attacks' performances over quantities referenced in Section \ref{secStrucTest}. Despite the varied uncertainty, APS-B continues to effectively balance perturbed observations with impact, and is able to identify high-quality but low-cardinality attacks. Likewise, as in the structural testing, the RME algorithm finds high-perturbation attacks of varied impact. In juxtaposition, the two R\&S approaches only identify the most impactful attack once out of the sixteen combinations. The APS-A algorithm performs much better across the uncertainty levels than the structural level; its impact is often among the top three of the examined attacks. Finally, no attack generates substantial impact on the path-attraction problem since the $\nicefrac{w_1}{w_2}$-value examined makes the cost of the attack less than its reward.

\begin{table}[H]
\centering
\caption{Mean Impact and Proportion of Observations Attacked (Uncertainty Testing)}
\label{tab:unc_attacks_allprobs}
\label{tab:unc_case}
\resizebox{0.8\textwidth}{!}{
\begin{tabular}{cccccccccc}
\hline
\multirow{2}{*}{Problem} & \multirow{2}{*}{Algorithm} & \multicolumn{2}{c}{Design Pt. 1}                    & \multicolumn{2}{c}{Design Pt. 2}                    & \multicolumn{2}{c}{Design Pt. 3}                    & \multicolumn{2}{c}{Design Pt. 4}                   \\ \cmidrule(lr){3-4} \cmidrule(lr){5-6}  \cmidrule(lr){7-8}  \cmidrule(lr){9-10} 
  &                  & Impact & $\nicefrac{\Delta}{|\mathcal{T}|}$ & Impact & $\nicefrac{\Delta}{|\mathcal{T}|}$ & Impact & $\nicefrac{\Delta}{|\mathcal{T}|}$ & Impact & $\nicefrac{\Delta}{|\mathcal{T}|}$ \\ \hline
  & APS-A                  & 0.255  & \textbf{0.145}                           & 0.254  & 0.140                           & 0.121  & \textbf{0.085}                           & 0.121  & 0.080                           \\ 
\multirow{2}{*}{State}& APS-B                  & \textbf{0.266}  & 0.150                           & 0.247  & 0.135                           & \textbf{0.133}  & 0.090                           & 0.127  & \textbf{0.085}                           \\ 
\multirow{2}{*}{Att.$^1$}& R\&S-A                    & 0.171  & 0.625                           & 0.162  & 0.550                           & 0.108  & 0.570                           & 0.116  & 0.625                           \\
& R\&S-B                    & 0.216  & 0.275                           & \textbf{0.259}  & 0.145                           & 0.122  & 0.345                           & \textbf{0.133}  & 0.090                           \\ 
& RME                     & 0.167  & 0.830                           & 0.156  & 0.810                           & 0.079  & 0.755                           & 0.077  & 0.765                           \\ \hline
& APS-A                  & 0.328  & \textbf{0.050}                           & 0.319  & \textbf{0.050}                          & 0.261  & \textbf{0.050}                           & 0.252  & \textbf{0.050}                           \\ 
\multirow{2}{*}{State}& APS-B                  & \textbf{0.329}  & \textbf{0.050}                           & \textbf{0.330}  & \textbf{0.050}                           & \textbf{0.266}  & \textbf{0.050}                           & 0.259  & \textbf{0.050}                           \\ 
\multirow{2}{*}{Rep.$^2$}& R\&S-A                    & 0.318  & 0.470                           & 0.322  & 0.570                           & 0.252  & 0.505                           & 0.249  & 0.560                           \\ 
&R\&S-B                      & 0.324  & 0.260                           & 0.327  & 0.220                           & 0.258  & 0.265                           & \textbf{0.266}  & 0.230                           \\ 
& RME                     & 0.313  & 0.755                           & 0.308  & 0.755                           & 0.246  & 0.725                           & 0.235  & 0.750                           \\ \hline
& APS-A                  & \textbf{2.171} & 0.075                           & \textbf{1.394} & 0.090                           & 1.629          & 0.075                           & \textbf{1.156} & 0.070                           \\ 
\multirow{2}{*}{Dist}& APS-B                  & 2.057          & \textbf{0.050}                  & 1.364          & \textbf{0.050}                  & 1.603          & \textbf{0.050}                  & 1.082          & \textbf{0.050}                  \\ 
\multirow{2}{*}{Disrupt.$^3$}& R\&S-A                    & 2.153          & 0.525                           & 1.364          & 0.595                           & 1.622          & 0.595                           & 1.071          & 0.615                           \\ 
& R\&S-B                     & 2.104          & 0.225                           & 1.352          & 0.185                           & \textbf{1.677} & 0.295                           & 1.089          & 0.185                           \\ 
& RME                     & 2.154          & 0.820                           & 1.387          & 0.900                           & 1.478          & 0.795                           & 1.104          & 0.900                           \\ \hline
& APS-A                  & \textbf{0.941}  & 0.045                           & \textbf{0.940}  & 0.105                           & \textbf{0.942}  & 0.065                           & \textbf{0.945}  & 0.060                           \\  
\multirow{2}{*}{Path} & APS-B                  & 0.950  & \textbf{0.000}                           & 0.950  & \textbf{0.000}                           & 0.950  & \textbf{0.000}                           & 0.950  & \textbf{0.000}                           \\  
\multirow{2}{*}{Att.$^4$} & R\&S-A                   & 0.942  & 0.365                           & 0.954  & 0.325                           & 0.955  & 0.090                           & 0.952  & 0.490                           \\  
& R\&S-B                     & 0.951  & 0.190                           & 0.964  & 0.295                           & 0.954  & 0.045                           & 0.955  & 0.145                           \\  
& RME                     & \textbf{0.941}  & 0.750                           & 0.953  & 0.750                           & 0.947  & 0.770                           & \textbf{0.945}  & 0.750                           \\ \hline
\multicolumn{10}{l}{\scriptsize $^1$ Impact is the difference between perturbed and unperturbed data of the probability of $i'$ at $t'$  } \\ [-1ex] 
\multicolumn{10}{l}{\scriptsize $^2$ Impact is the difference between unperturbed and perturbed data of the probability of $i'$ at $t'$  } \\ [-1ex]
\multicolumn{10}{l}{\scriptsize $^3$ Impact is the KL divergence between state distributions under unperturbed and perturbed data} \\ [-1ex] 
\multicolumn{10}{l}{\scriptsize $^4$ Impact is the Normalized-Hamming distance of most-likely state sequences perturbed data and attacker's goal} 
  \end{tabular}
  }
\end{table}

The overall performance of the attacks appears to be influenced most by $\lambda$, i.e., the probability of an attack changing the true observation, in the state-attraction and -repulsion problems. However, this pattern does not necessarily hold in the other problems. Within the distribution-disruption instances, $\kappa$ has the most influence, whereas both $\lambda$ and $\kappa$ have a marginal to nil effect on the path-attraction instances as a consequence of the high $\nicefrac{w_1}{w_2}$-values. The APS-A, APS-B and R\&S-B algorithms identify a high-quality attack in the state-attraction, state-repulsion, and distribution-disruption problems. However, the identified attacks balance impact and data perturbation differently between algorithms. The totality of these results illustrate that, in general, the trends identified in the previous sub-section with regard to algorithmic performance are valid with varied uncertainty levels while the improved performance of APS-A is an exception.

\subsection{Case Study: Attacking an HMM for Part-of-Speech Tagging} \label{sec:CaseStudy}

To highlight the practical relevance of our methods, we consider attacks against a larger-scale HMM used for natural-language processing (NLP). In particular, we are motivated by part-of-speech (POS) tagging, a widely used approach in the lexical analysis of text data. POS tagging improves accuracy of text analysis by reducing the computational effort required for data processing and revealing the syntactic structure of sentences. 
Herein, hidden states correspond to a POS, and the observations are words. Emission probabilities refer to the probability of a word given a POS, whereas transition probabilities capture POS sequencing. Attacks in this section are accomplished using an Apple M2 workstation with 8 GB of RAM and 8 CPU cores in order to demonstrate algorithm efficacy even by using a standard laptop.

Concretely, we train an HMM on the Named-Entity-Recognition dataset \citep{kaggle2023}. The top-300 words in the database are considered, along with all 29 POSs, in the examination of 30-word text strings. The trained HMM is taken as the decision-maker's model having $|\mathcal{Q}|=29$, $|\mathcal{X}|=300 $, $|\mathcal{T}|=30$. As in previous sections, the attacker's prior beliefs are centered about the decision maker's true model. We work in a low uncertainty scenario. We implement R$\&$S-B method for this case-study due to its better scalability with respect to $|\mathcal{X}|$.

Let the following pre-processed phrase from the NER data be the uncorrupted text:

\begin{quote}
    \textit{mr. said among countries six party talks (among china, south korea, japan, russia, united states north korea) north korea's nuclear program}
\end{quote}

\begin{figure}[htb]
\begin{subfigure}[b]{0.45\textwidth}
    \centering
    \includegraphics[width=0.9\textwidth]{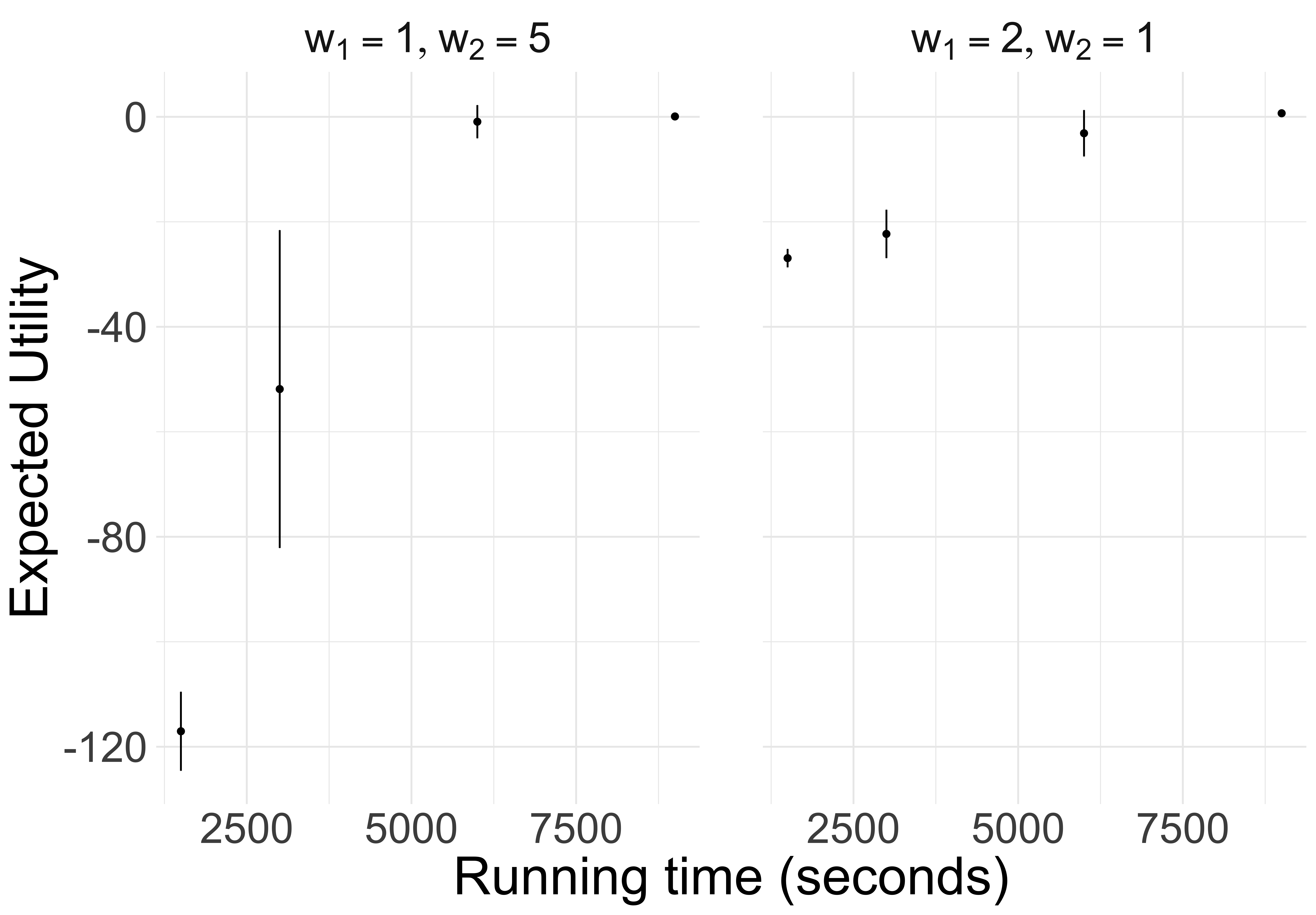}
    \caption{State-Attraction}
    \label{fig:att}
\end{subfigure} 
\begin{subfigure}[b]{0.45\textwidth}
    \centering
    \includegraphics[width=0.9\textwidth]{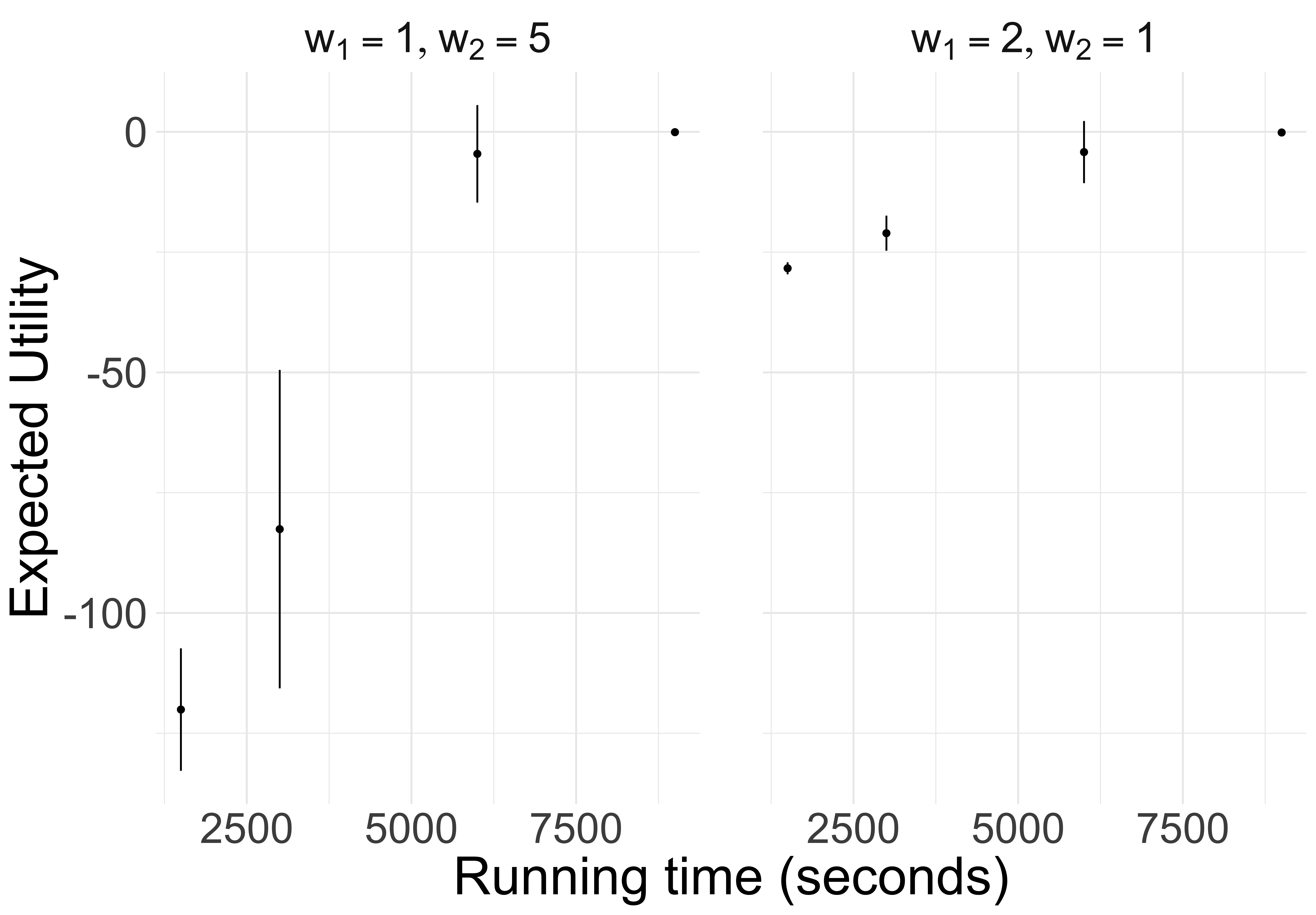}
    \caption{State-Repulsion}
    \label{fig:rep}
\end{subfigure} 
\\
\begin{subfigure}[b]{0.45\textwidth}
    \centering
    \includegraphics[width=0.9\textwidth]{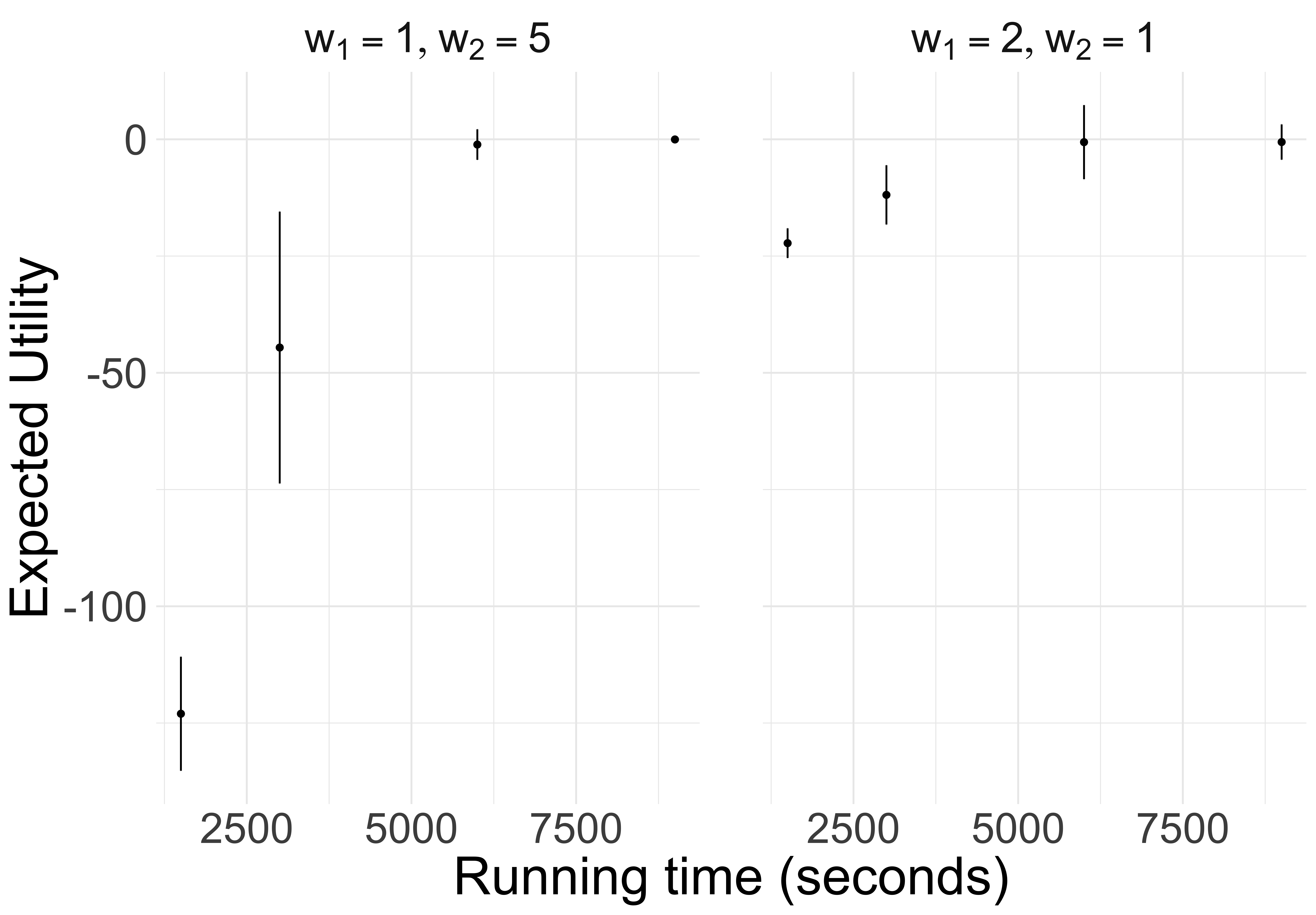}
    \caption{Distribution Disruption}
    \label{fig:disr}
\end{subfigure} 
\begin{subfigure}[b]{0.45\textwidth}
    \centering
    \includegraphics[width=0.9\textwidth]{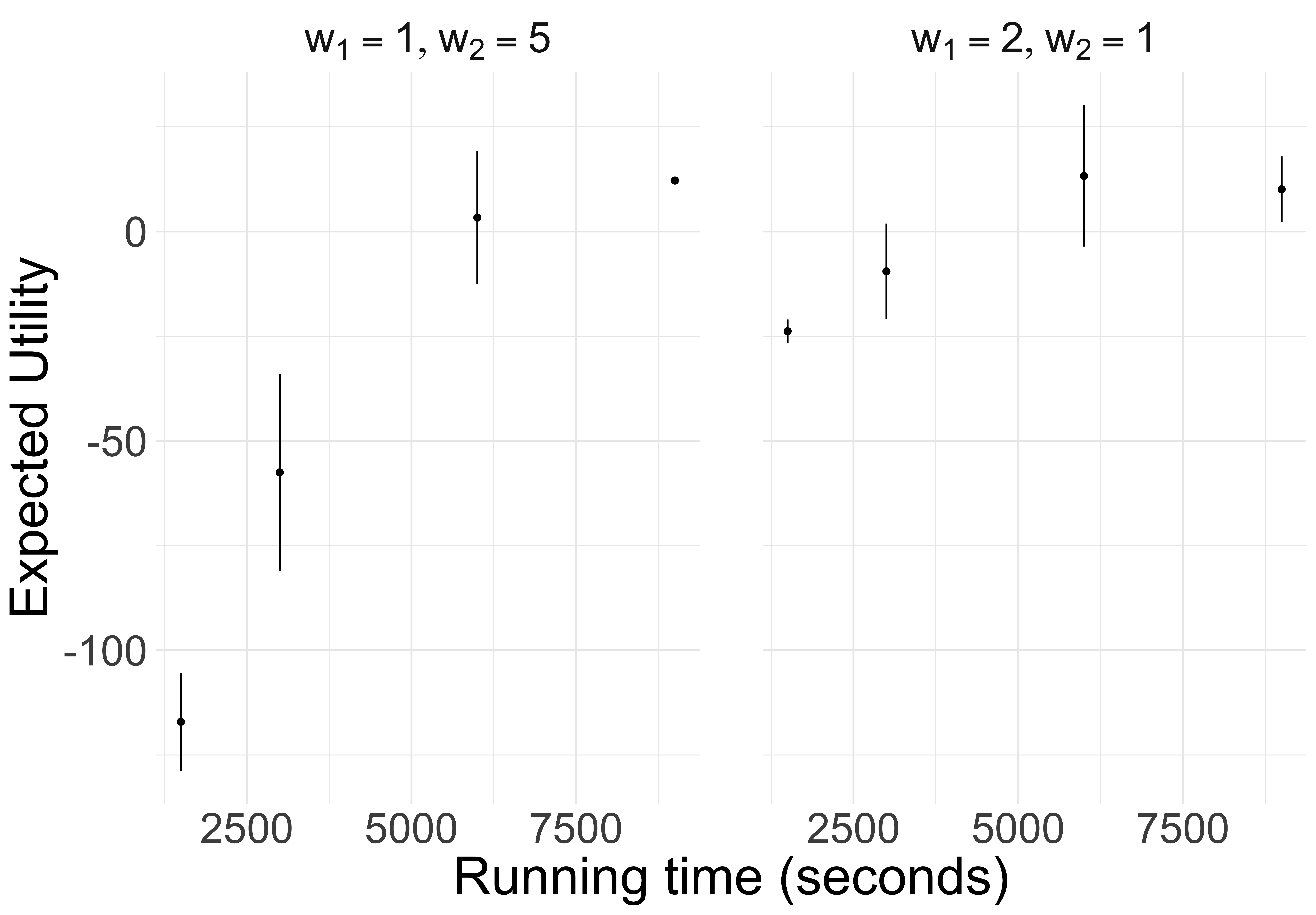}
    \caption{Path Attraction}
    \label{fig:path}
\end{subfigure} 
\centering
\caption{\centering Expected Utility of 
Attacks Plus/Minus One Standard Deviation in the POS Case Study}
\label{fig:casestudy}
\end{figure}

\noindent Figure \ref{fig:casestudy} examines the utility of attacking this phrase via the R$\&$S-B method across the state-attraction, state-repulsion, distribution-disruption and path-attraction problems. Specific details regarding the associated POS information for the state-attraction and -repulsion problems, as well as the desired POS sequences for the path-attraction problem are available online in our code repository. Attack success probabilities are similar to the previous sections and are also detailed in the code repository. For each of the problems, we again vary the objective weights. We alternatively set $(w_1,w_2)$ to $(1,5)$ and $(2,1)$ to explore its effect. These weights are selected to ensure problem difficulty; namely, we do not want the attacker to be incentivized to make drastic changes to the uncorrupted text. For each of the problem-and-weight combinations, the attack was allowed to search the solution space for 1500, 3000, 6000 and 9000 seconds. Each experiment was performed 10 times to enable estimation of the mean expected utility as well as its standard deviation.

Examination of Figure \ref{fig:casestudy} reveals several noteworthy insights. For each of the ten R$\&$S-B attacks performed on each problem-and-weight pair, a different $\hat{z}^*$ may be identified based on the stochasticity inherent in the simulated-annealing optimization. Inspection of Figure \ref{fig:casestudy} confirms intuition that lower computation times result in attacks of lesser quality. As additional computation time is allocated, it is apparent that not only does the solution quality improve, but the reduction in variability suggest that the algorithms converge toward an attack across the ten runs. 
Alternatively, variability in the expected utility measurements is reduced when $(w_1,w_2)=(2,1)$, implying that, even when allocated less computation time, the runs converge toward comparable solutions.

Additional information regarding the performance of the superlative attacks generated using 9000 seconds of computational effort is detailed in Table \ref{tab:casestudy}. The same performance metrics are tallied as in Table \ref{tab:unc_attacks_allprobs}. Notably, whenever $w_2>w_1$, it can be observed that the attack is minimally perturbing the true data with limited impact due to the higher weighted cost. More aggressive and impactful attacks are identified when $w_1>w_2$. The totality of these results highlight that a conventional machine can effectively corrupt HMMs in a grey-box setting, even when the optimal attack minimally perturbs the uncorrupted data.

\begin{table}[htbp!]
\centering
\caption{Mean Impact, Proportion of Observations Attacked and Expected Utility (NER Experiment)} 
\label{tab:casestudy}
\resizebox{\textwidth}{!}{
\begin{tabular}{ p{0.30\textwidth}p{0.08\textwidth}p{0.08\textwidth}p{0.16\textwidth}p{0.16\textwidth}p{0.22\textwidth}}
\hline
Problem & $w_1$ & $w_2$ & Impact & $ \nicefrac{\Delta}{|T|}$ & Expected Utility \\ \hline
\multirow{ 2}{*}{State Attraction$^{1}$}  & 1 & 5 & 0.00 & 0.00 & 0.07 \\
 & 2 & 1 & 0.77 & 0.13 & 0.68\\ \hline
\multirow{ 2}{*}{State Repulsion$^{2}$}  & 1 & 5 & 0.00 & 0.00 & -0.07\\
& 2 & 1 & 0.00 & 1.0 & -0.14 \\ \hline
\multirow{ 2}{*}{Distribution Disruption$^{3}$} & 1 & 5 & 1.47 & 0.01 & -0.03  \\ 
& 2 & 1 & 11.21 & 0.77 & -0.59\\ \hline
\multirow{ 2}{*}{Path Attraction$^4$}  & 1 & 5 & 0.53 & 0.00 & 12.17\\
& 2 & 1 & 0.71 & 0.74 & 10.08\\
\hline
\multicolumn{6}{l}{\scriptsize $^1$ Impact is the difference between perturbed and unperturbed data of the probability of $i'$ at $t'$  } \\ [-1ex] 
\multicolumn{6}{l}{\scriptsize $^2$ Impact is the difference between unperturbed and perturbed data of the probability of $i'$ at $t'$  } \\ [-1ex]
\multicolumn{6}{l}{\scriptsize $^3$ Impact is the KL divergence between state distributions under unperturbed and perturbed data} \\ [-1ex] 
\multicolumn{6}{l}{\scriptsize $^4$ Impact is the Normalized-Hamming distance of most-likely state sequences perturbed data and attacker's goal} \\ [-1ex] 
\end{tabular}
  }
\end{table}

This feature is exemplified in the state-attraction case having $(w_1,w_2)=(2,1)$ wherein the attacker wishes to encourage the classification of the fifth word as a singular proper noun. The superlative attack identified across all ten runs is listed below:

\begin{quote}
\textit{mr. said among countries u.s. party talks (among china, south korea, japan, russia, united states north korea) north korea's nuclear program  }
\end{quote}

\noindent Notably, this text string is quite similar to the uncorrupted data. Their variations could easily be attributed to a typographical error or a bug in the text's pre-processing routine. However, this small change greatly increases the probability that the attacker achieves his goal; the attack increases the posterior probability of a singular-proper-noun by 0.77. The ability of this attack to thwart the decision maker's HMM with minimal changes to the true data is reminiscent of adversarial examples in other applications, e.g., computer vision \citep{goodfellow2014explaining}.

\section{Conclusion} \label{secConc}

Probabilistic graphical models are fundamental to modern technology \citep{koller2009probabilistic}. Diverse applications, from natural language processing to autonomous navigation, have all leveraged such models to achieve state-of-the-art results. Within this manuscript, we have illustrated that, like other machine learning methodologies, these models are susceptible to attack. 

Focusing on HMMs, dynamic Bayesian networks with a specific structural form, we have formulated a suite of corruption problems for filtering, smoothing and decoding inferences. Leveraging an ARA perspective, a collection of general solution methods was also developed by alternatively viewing the problem from frequentist and Bayesian perspectives. Extensive empirical testing on these algorithms illustrated the devastating impacts of even minor data perturbations on HMM inferences. Moreover, this testing also examined the effects of HMM structure and uncertainty on the developed algorithms highlighting that, even for more complex instances, one may identify high-quality attacks in a reasonable amount of time. Our case study highlighted the real-world applicability of our attacks, implying that corrupted HMM models can create adversarial HMM examples akin to those that thwart computer vision algorithms \citep{goodfellow2014explaining}.

However, the development of these attacks begets numerous avenues of future inquiry. For example, given the vulnerability of HMMs to data perturbation, there is an obvious need for more robust HMM inference algorithms. The efficacy of the developed attacks on a trained HMM also suggests that traditional approaches to HMM learning may be vulnerable. Future research should therefore focus on the corruption of the Baum-Welch algorithm as well as on modifications to robustify it. The same variety of questions can also be formulated with respect to other probabilistic graphical models (e.g., Latent Dirichlet Allocation models). Notably, the problems and attacks provided herein can be directly extended to inference over alternative dynamic Bayesian networks (e.g., Kalman filters, autoregressive HMMs). Nevertheless, AML research upon standard HMMs is by no means exhausted either; varied assumptions developed herein can be relaxed to distinct, real-world settings (e.g., real-time inference or uncertain $\mathcal{Q}$ and $\mathcal{X}$). Moreover, given the generality of the HMM corruption problems, the variety of prior distributions available to the attacker, and the flexible hyperparameterization of the R\&S heuristic, additional experimentation on our algorithms is a worthwhile endeavor, as is the development of alternative attacks. Therefore, while this research has incrementally developed AML techniques for probabilistic graphical models, avenues of future inquiry abound.

\section*{Acknowledgments}

\begin{wrapfigure}{l}{0.25\textwidth}
    \centering
    \includegraphics[width=0.25\textwidth]{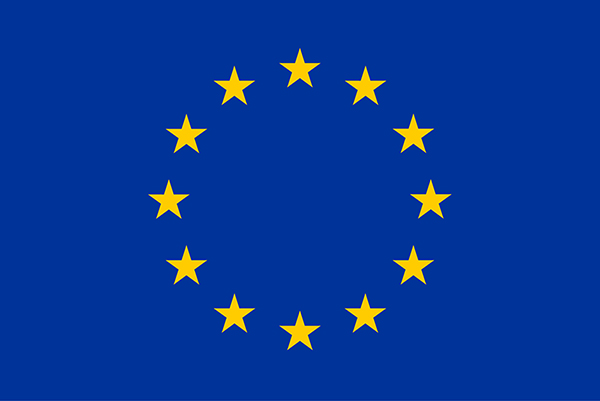}
\end{wrapfigure}

This work is supported by Air Force Scientific Office of Research (AFOSR), USA award FA-9550-21-1-0239; AFOSR European Office of Aerospace Research and Development award FA8655-21-1-7042, EOARD-AFOSR project RC2APD grant 13324227, and the Spanish Ministry of Science and Innovation program PID2021-124662OB-I00. Any opinions, findings, and conclusions or recommendations expressed are those of the authors and do not necessarily reflect the views of the sponsors. T.E. is also supported by Jerry Fields Endowed Chair award and Steven R. “Steve” Gregg Excellence Endowed Professorship. J.M.C. is supported by a fellowship from ”la Caixa” Foundation (ID 100010434), whose code is LCF/BQ/DI21/11860063. This project has received funding from the  European Union's Horizon 2020 Research and Innovation Programme under Grant Agreement No. 101021797.

\appendix
\section{Implementation Details of the R\&S Heuristic} \label{appendix_RandS}

This appendix provides implementation details of the R\&S heuristics utilized in Sections \ref{secStrucTest} and \ref{secUncTest}. We discuss the functional approximation $\hat{\mu}(z^n|\theta^n)$, the associated state variables (i.e., $\theta^n$), the system model (i.e., $S^M$), and the optimization routines used to select $z^n$ given some parameterization of $\hat{\mu}(z^n|\theta^n)$. Associated code implementing the procedures discussed herein is available at https://github.com/roinaveiro/corrupting\_hmms.

As a functional approximation of an attack's value, we use a fully connected neural network. Different architectures were tested and the superlative was selected, i.e., see \ref{appendix_hyper_selection}.
The system state at each iteration $n$ captures the current parameterization of this neural network, e.g., the arc weights. The updated state variables (i.e., $\theta^{n+1})$ are identified using a stochastic-gradient-descent algorithm as the system model. In particular, the Adam optimizer is utilized. The performance of different learning rates is tested as well; see  \ref{appendix_hyper_selection} for additional details.

To iteratively select $z^n$, the R\&S heuristic uses an $\varepsilon$-greedy policy to encourage exploration. Within each iteration of R\&S, the greedy action that maximizes the objective function under our current belief state is of foremost interest. 
This requires solving a non-linear, integer optimization problem. Unfortunately, given the high cardinality of the attack space (i.e., $\mathcal{Z}$),
solving this optimization via complete ennumeration is infeasible; other canonical branch-and-bound techniques are likely to struggle as well. 
Therefore, we compare the performance of two meta-heuristics to approximate this solution: Monte Carlo Tree Search (MCTS) and Simulated Annealing. Implementation details for both algorithms are provided subsequently in \ref{appendix_MCTS} and \ref{appendix_SA}. Hyperparameter tuning is discussed in \ref{appendix_hyper_selection}.

\subsection{Monte Carlo Tree Search}\label{appendix_MCTS}

To find the greedy action using MCTS, we first reformulate this optimization problem as an undiscounted, sequential-decision problem wherein the greedy attack is constructed iteratively by time period. 
With slight recycling of notation, this problem is formally defined by the tuple $(\mathcal{S}, \mathcal{A}, \mathcal{E}, \mathcal{R})$ wherein $\mathcal{S}$ is the set of all partially and fully filled attack vectors, $\mathcal{A}$ is the set of feasible actions at each $s \in \mathcal{S}$, $\mathcal{E}$ is a map from $\mathcal{S} \to \mathcal{S}$, and $\mathcal{R}$ is the set of instantaneous rewards at each  $s \in \mathcal{S}$. At the beginning of each episode, we start with an empty attack vector (i.e., $s_0$). An episode concludes after the attack vector is fully specified. This is accomplished by selecting an emission for each time $t$ from $\mathcal{A}(s_t)$ whereby $s_t$ captures all emissions inserted into the attack up to this time. The transition function $\mathcal{E}$ deterministically maps a state-action pair $(s_t, a_t)$ into the next state $s_{t+1}$ by inserting the selected emission into the attack vector. The set of state rewards, $\mathcal{R}$, varies depending upon whether the state is complete or incomplete, i.e., whether emissions have or have not been selected for every $t \in \mathcal{T}$, respectively. The reward associated with any incomplete state is zero. However, the reward for complete states (i.e., fully specified attack vectors) corresponds to the evaluation of $\hat{\mu}(s|\theta^n)$.

Using this foundation, our MCTS algorithm constructs a tree comprised of all possible attack vectors. Its nodes correspond to some $\mathcal{S}' \subseteq \mathcal{S}$. At each iteration of the algorithm, the tree is traversed by selecting actions via a tree policy until a node with a child outside of the partially constructed tree is reached. If such a node is reached, actions are henceforth selected at random until a complete state is achieved. After reaching a complete state, its reward is computed and used to update the value of nodes that have been traversed in the tree such that better nodes are more likely to be selected in the future.

More specifically, given a partially explored tree $\mathcal{G}$, each MCTS iteration consists of four distinct steps:
\begin{enumerate}
    \item \textit{Selection}. Starting from the root node,  (i.e., the empty attack vector, $s_0$), actions are selected using the tree policy until a node with some child outside the tree is reached. In our implementation, we use the UCT selection criterion as the tree policy's basis. That is, at state $s_t$, action $a_{t}$ (leading to node $\mathcal{E}(s_t, a_t)$) is identified using
    \begin{equation*}
        \pi^{\mathcal{G}}(s_t) = \argmax_{a \in \mathcal{A}(s_t)} Q(s_t, a) + c \cdot  \sqrt{\frac{2 \log N(s_t)}{N(s_t, a)}}
    \end{equation*}
    where $Q(s_t, a)$ is a Monte Carlo estimate of the state-action value, $N(s_t)$ is the number of visits to the parent node, $N(s_t, a)$ is the number of times action $a$ has been taken at node $s_t$ and $c$ is an exploration parameter. In our implementation, we fix $c=0.5$.
    
    \item \textit{Expansion}. If a leaf node is reached in the selection step, it is appended to the tree and the simulation step begins.
    
    \item \textit{Simulation}. Actions outside the tree are selected uniformly at random from the available actions until a complete state has been reached.
    
    \item \textit{Backpropagation}. Upon reaching a complete state, its reward $R$ is evaluated by invoking $\hat{\mu}(s|\theta^n)$. This is used to update the value of the leaf node reached in the selection step as well as each of its parents. For every $(s',a)$ along the traversed path in the tree, the following updates are performed:

    \begin{eqnarray*}
        N(s', a) &\leftarrow& N(s',a) + 1\\
        Q(s',a)  &\leftarrow& Q(s',a) + \frac{R - Q(s',a)}{N(s',a)}
    \end{eqnarray*}
    
\end{enumerate}

\noindent In our implementation of the R\&S heuristic, MCTS is used to approximate the greedy policy in each iteration of R\&S (i.e., each time Step \ref{stepIDaction} is reached in Algorithm \ref{algRandS}). In particular, we perform $1,000$ MCTS iterations and record the best complete state reached as the approximate greedy action.

\subsection{Simulated Annealing} \label{appendix_SA}

The second meta-heuristic used to find the action maximizing $\hat{\mu}(z^n|\theta^n)$ is simulated annealing (SA) \citep{KGV:1983}. SA generates a Markov chain in the space of possible attacks, whose stationary distribution is proportional to $\exp(\hat{\mu}(z^n|\theta^n) / T_e)$, where $T_e$ is gradually decreased to 0 according to some pre-specified annealing schedule. We use Gibbs sampling to generate such a Markov chain, sequentially sampling from full conditional of the stationary distribution for each $z^n_t$ conditioned on $z^n_{-t}$. $T_e$ is reduced according to the exponential decay annealing schedule suggested by \cite{spears1993simulated}. Therefore, at the $j^{th}$ iteration of SA, we have $T_e = \exp\left( -l\cdot j / \mathcal{T}  \right)$, where $l$ is set to 5.

\subsection{Hyperparameter Tuning} 
\label{appendix_hyper_selection}

This appendix explains how hyperparameter tuning was performed for the two R\&S variants. 
Specifically, we varied the following hyperparameters: the architecture of the multilayer perceptron (MLP) neural network (i.e., the number of neurons in the two layers), the number of  SA or MCTS iterations used to find the greedy action, the learning rate, and $\varepsilon$. Tables \ref{tab:rs_mcts_hyp} and \ref{tab:rs_sa_hyp} lists the tested hyperparameter combinations.  

\begin{table}[H]
\centering
\resizebox{0.6\textwidth}{!}{
\begin{tabular}{ccccc}
\hline
Combination      &   MLP Architecture & Iterations & Learning rate & $\varepsilon$ \\ \hline
1                &   \{16; 8\}           &       100     &         0.005      &  0.05          \\ 
2                &  \{32; 16\}             &        10    &          0.005     &   0.05         \\ 
3                &   \{16; 8\}           &     100       &        0.1       &  0.005          \\ 
4                &    \{32; 16\}          &       10     &     0.1          &    0.005        \\ 
5 &  \{64; 64\}            &    100        &          0.005     &    0.05         \\ 
6 & \{64; 64\}             &        100    &          0.1     & 0.05           \\ \hline
\end{tabular}
}
\caption{Hyperparameter Combinations for the R$\&$S-A Algorithm.}
\label{tab:rs_mcts_hyp}
\end{table}

\begin{table}[H]
\centering
\resizebox{0.6\textwidth}{!}{
\begin{tabular}{ccccc}
\hline
Combination      & MLP Architecture & Iterations & Learning rate & $\varepsilon$ \\ \hline
1                &   \{16; 8\} &        50    &        0.005       &   0.05         \\
2                &  \{32; 16\}  &        10    &       0.005     &  0.05          \\ 
3                &   \{16; 8\}  &        50    &       0.1        &  0.005          \\ 
4                &   \{32; 16\}  &        10    &      0.1     &  0.005          \\ 
5 & \{64; 64\}      &   100   &      0.005     &    0.05           \\ 
6 & \{64; 64\}       &   100     &      0.1         &  0.05          \\  \hline
\end{tabular}
}
\caption{Hyperparameter Combinations for the R$\&$S-B Algorithm.}
\label{tab:rs_sa_hyp}
\end{table}

An extensive grid search on the hyperparameters was not performed; instead, a simple search procedure was conducted to (1) examine how the algorithms behave under different hyperparameters and (2) identify settings that perform well on the examined instances. To do so, we replicated the testing procedure described in Sections \ref{secStrucTest} and \ref{secUncTest} for each combination. Each algorithm-and-hyperparameter combination was tested 10 times for run times of $\{15,30,45,...,1200\}$ seconds on each problem instance. Mean expected utility values plus/minus two standard deviations were recorded for each run time across all 10 repetitions. 
Algorithm performance was judged by their convergence time, solution quality, and relative balance of these characteristics. A qualitative assessment of these measures was utilized for model selection. 

Tables \ref{tab:res_hyp_hmm_setup} and \ref{tab:best_hyp_unc_exp} indicate the superlative hyperparameter combinations for each problem-and-design-point pair. The indicated hyperparameter setting is leveraged within Sections \ref{secStrucTest}  and \ref{secUncTest}, respectively.  Notably, no single combination is dominant across all instances, but some combinations were always dominated by another. For example, combination four and three of the R\&S-A and -B algorithms, respectively, are dominated in the Section \ref{secStrucTest} instances. Likewise, combination six is dominated across all problem-and-design-point pairs. Analogous patterns are apparent in the Section \ref{secUncTest} testing as well. 

\begin{table}[H]
\centering
\caption{Superlative Hyperparameter Combination for Section Structural Testing}
\label{tab:res_hyp_hmm_setup}
\resizebox{0.5\textwidth}{!}{
\begin{tabular}{cccccc}
\hline
\multirow{2}{*}{Problem}                                                           & \multirow{2}{*}{Algorithm} & \multicolumn{4}{c}{Design Point} \\ \cline{3-6} 
                                                                                   &                         & 1      & 2      & 3     & 4     \\ \hline
\multirow{2}{*}{\begin{tabular}[c]{@{}c@{}}State-Attraction\end{tabular}}       &  R\&S-A                      & 2    & 3     &  2   &  2  \\
                                                                                   &  R\&S-B                    & 2     & 1     & 2    & 2    \\ \hline
\multirow{2}{*}{\begin{tabular}[c]{@{}c@{}}State-Repulsion\end{tabular}}         &  R\&S-A                      & 2      &  3     &  2    &   2   \\
                                                                                   &  R\&S-B                    &  2     & 1    & 2     &  5     \\ \hline
\multirow{2}{*}{\begin{tabular}[c]{@{}c@{}}Distribution-Disruption\end{tabular}} &  R\&S-A                      &  2     &    5    &    2  &  1  \\
                                                                                   &  R\&S-B                    &   2    &    5   &  2   &  5    \\ \hline
\multirow{2}{*}{\begin{tabular}[c]{@{}c@{}}Path-Attraction\end{tabular}}      &  R\&S-A                      &   2    &  1   &  1 &  1 \\
                                                                                   &  R\&S-B                    &  4      &    1   &  2    &   2  \\ \hline
\end{tabular}
}
\end{table}

\begin{table}[H]
\centering
\caption{Superlative Hyperparameter Combination for HMM Uncertainty Testing}
\label{tab:best_hyp_unc_exp}
\resizebox{0.5\textwidth}{!}{
\begin{tabular}{cccccc}
\hline
\multirow{2}{*}{Problem}                                                           & \multirow{2}{*}{Algorithm} & \multicolumn{4}{c}{Design Point} \\ \cline{3-6} 
                                                                                   &                         & 1      & 2      & 3     & 4     \\ \hline
\multirow{2}{*}{\begin{tabular}[c]{@{}c@{}}State-Attraction\end{tabular}}       &  R\&S-A                      &   3     &  3     & 5     & 5     \\
                                                                                   &  R\&S-B                   & 5      & 1      &    2   & 1      \\ \hline
\multirow{2}{*}{\begin{tabular}[c]{@{}c@{}}State-Repulsion\end{tabular}}         &  R\&S-A                      & 5      & 5      & 5     & 5     \\
                                                                                   &  R\&S-B                    & 2       & 2       & 2      & 2      \\ \hline
\multirow{2}{*}{\begin{tabular}[c]{@{}c@{}}Distribution-Disruption\end{tabular}} &  R\&S-A                      & 5      & 2      & 5     & 2     \\
                                                                                   &  R\&S-B                    & 2       & 2      & 2      & 2     \\ \hline
\multirow{2}{*}{\begin{tabular}[c]{@{}c@{}}Path-Attraction \end{tabular}}      &  R\&S-A                      & 2       & 2       & 1      & 1      \\
                                                                                   &  R\&S-B                    & 4      & 4      & 4     & 4     \\ \hline
\end{tabular}
}
\end{table}

\bibliography{aps_biblio}

\end{document}